\documentclass [11pt]{article}

\usepackage{graphicx}
\usepackage{amsmath,amssymb} 
\usepackage{latexsym}
\usepackage{mathrsfs}
\usepackage{color}
\usepackage{axodraw}
\definecolor{red}{rgb}{1,0,0}

\makeatletter
\def\section{\@startsection {section}{1}{\z@}{-3.5ex plus -1ex minus
 -.2ex}{2.3ex plus .2ex}{\large\bf}}
\def\subsection{\@startsection{subsection}{2}{\z@}{-3.25ex plus -1ex
minus -.2ex}{1.5ex plus .2ex}{\normalsize\bf}}
\makeatother
\makeatletter

\@addtoreset{equation}{section}

\makeatother
\def\bea{\begin{eqnarray}} \def\eea{\end{eqnarray}}
\def\be{\begin{equation}} \def\ee{\end{equation}} \def\nn{\nonumber}
  \def\Z{{\bf Z}}

\newcommand{\promille}{%
  \relax\ifmmode\promillezeichen
        \else\leavevmode\(\mathsurround=0pt\promillezeichen\)\fi}
\newcommand{\promillezeichen}{%
  \kern-.05em%
  \raise.5ex\hbox{\the\scriptfont0 0}%
  \kern-.15em/\kern-.15em%
  \lower.25ex\hbox{\the\scriptfont0 00}}

\newcommand{\ol}{\overline}
\newcommand{\wt}{\tilde}
\newcommand{\bs}{\boldsymbol}

\begin{document}

\thispagestyle{empty}

\begin{center}

\hfill SISSA-27/2011/EP \\
\hfill DFPD-2011/TH/08\\

\begin{center}

\vspace*{0.5cm}

{\Large\bf Leptons in Holographic Composite Higgs Models\\[2mm] with Non-Abelian Discrete Symmetries}

\end{center}

\vspace{1.4cm}

{\bf Claudia Hagedorn$^{a}$ and Marco
Serone$^{b,c}$}\\

\vspace{1.2cm}

${}^a\!\!$
{\em Dipartimento di Fisica `G.~Galilei', Universit\`a di Padova
\\
INFN, Sezione di Padova, Via Marzolo~8, I-35131 Padua, Italy}

\vspace{.3cm}

${}^b\!\!$
{\em SISSA and INFN, Via Bonomea 265, I-34136 Trieste, Italy} 

\vspace{.3cm}

${}^c\!\!$
{\em ICTP, Strada Costiera 11, I-34151 Trieste, Italy}

\end{center}

\vspace{0.8cm}

\centerline{\bf Abstract}
\vspace{2 mm}
\begin{quote}

We study leptons in holographic composite Higgs models, namely in models possibly admitting a weakly coupled description
in terms of five-dimensional (5D) theories. We introduce two scenarios leading to Majorana or Dirac neutrinos, 
based on the non-abelian discrete group $S_4\times \Z_3$ which is responsible for nearly tri-bimaximal  lepton mixing. 
The smallness of neutrino masses is naturally explained and normal/inverted mass ordering can be accommodated.
We analyze two specific 5D gauge-Higgs unification models in warped space as concrete examples of our framework.
Both models pass the current bounds on Lepton Flavour Violation (LFV) processes.
We pay special attention to the effect of so called boundary kinetic terms that are the dominant source of LFV.
The model with Majorana neutrinos is compatible with a Kaluza-Klein vector mass scale $m_{KK}\gtrsim 3.5$ TeV,  
which is roughly the lowest scale allowed by electroweak considerations. 
The model with Dirac neutrinos, although not strongly constrained by LFV processes and data on lepton mixing, suffers from a too large deviation of the neutrino coupling to the $Z$ boson 
from its Standard Model value, pushing $m_{KK}\gtrsim  10$ TeV.  

\end{quote}

\vfill

\newpage

\section{Introduction}

The idea that the Standard Model (SM) Higgs might be a composite particle arising from a strongly coupled theory \cite{Kaplan} has received considerable attention lately. One of the main reasons of this renewed interest comes from the observation that the composite Higgs paradigm is closely related to theories in extra dimensions \cite{ArkaniHamed:2000ds}. This connection is particularly transparent in Randall-Sundrum (RS) models \cite{Randall:1999ee}, thanks to the AdS/CFT duality \cite{Ads-cft}. More precisely, certain theories in extra dimensions, including RS models, can be seen as a (relatively) weakly coupled description of a sub-set of 4D composite Higgs models.  They consist of two sectors:
an ``elementary'' sector, which includes the gauge and fermion fields of the SM, and a ``composite'' sector, which is strongly coupled and gives rise to the SM Higgs. The form of the couplings between these two sectors is not the most general one allowed by symmetry considerations only, but is more constrained. We denote in the following this more constrained class of models as
Holographic Composite Higgs Models (HCHM).

The flavour structure of HCHM has been studied in detail in the past mostly in the 5D context of RS models  with fermion and gauge fields in the bulk and it has been shown to be particularly successful \cite{flavor}. It automatically implements the idea of \cite{ArkaniHamed:1999dc} to explain the hierarchy of the quark and charged lepton masses in terms of field localization in an extra dimension. Moreover, HCHM are equipped with a built-in GIM mechanism that goes under the name of RS-GIM \cite{Agashe:2004cp} and automatically protects the SM fields from possibly large flavour violating interactions coming from the composite sector.\footnote{Despite this protection mechanism, a CP violation problem is still present in the quark sector \cite{Agashe:2004cp,CFW}.} 

Small neutrino masses and large lepton mixing are not easily accommodated in this set-up, because the large mixing potentially leads to excessive LFV.
Neutrino oscillation experiments clearly show that the Pontecorvo-Maki-Nakagawa-Sakata (PMNS) mixing matrix has a very peculiar structure well compatible with Tri-Bimaximal (TB)  mixing \cite{HPS}. 
There has been much progress in recent years in explaining TB lepton mixing and the absence of LFV 
interactions for charged leptons by means of discrete non-abelian symmetries. 
It is thus natural to apply such symmetries also in the context of HCHM in order to resolve the aforementioned problems. 

Aim of this paper is to introduce a class of HCHM where, thanks to a non-abelian discrete symmetry, lepton mixing is nearly TB, and at the same time bounds on LFV processes in the charged lepton sector are satisfied (see \cite{Agashe:2008fe} for other proposals). The mass spectrum in the neutrino sector can be normally or inversely ordered. The pattern of flavour symmetry breaking is dictated by symmetry considerations only, without relying on extra assumptions \cite{Fitzpatrick:2007sa} or specific mechanisms for the breaking of the flavour symmetry, such  as the ones used in \cite{Csaki,delAguila:2010vg} (see also \cite{Kadosh:2010rm}) in the case of $A_4$ to reproduce TB mixing \cite{A4notenough}. We discuss the case of flavour symmetry breaking in the elementary and composite sectors to certain non-trivial subgroups of the original symmetry {\it without}  advocating an explicit realization of the breaking.\footnote{For 5D models, this is the flavour counterpart of breaking by boundary conditions, commonly used for gauge symmetry breaking.} In particular, no flavons or other specific sources of flavour breaking are present in our set-up. We consider in this paper the discrete group $S_4 \times \Z_3$. The group $S_4$ has been shown \cite{S4Lam} to be the minimal group giving rise to TB lepton mixing using symmetry principles only. The presence of an irreducible two-dimensional representation is another feature of $S_4$. Such a representation allows to disentangle the symmetry properties of the third generation from the first two and is expected to be important when applying the flavour symmetry in the quark sector. 

We focus on two possible scenarios which only differ in the way SM neutrinos get a mass.
In the first one, the SM neutrinos are Majorana fermions and the type I see-saw mechanism explains the smallness of their masses, with no need to introduce additional (intermediate) mass scales in the theory. 
In the second one, SM neutrinos are Dirac fermions and tiny Yukawa couplings are naturally explained by the ultra-composite nature of the right-handed (RH) neutrinos \cite{ArkaniHamed:1999za}. 
In both scenarios, the flavour symmetry is broken to $\Z_2 \times \Z_2 \times \Z_3$ in the elementary and to $\Z_3^{(D)}$ in the composite sector. Note that the strength of this symmetry breaking is in general expected to be $\mathcal{O}(1)$.
In the composite sector for the charged leptons such a large breaking is actually favoured, because it allows to decrease the degree of compositeness of SM  leptons, suppressing large deviations from the SM $Z\tau\bar\tau $ coupling.\footnote{ This is an important point, also because uncalculable contributions to LFV processes (and flavour preserving quantities as well)  coming from higher dimensional operators are sub-leading with respect to the calculable ones only if the SM fields are sufficiently elementary.} The breaking felt by neutrinos in the composite sector is instead required to be weak in the Majorana scenario, in order to not perturb too much TB lepton mixing. 
An alternative is to resort to an extra symmetry protecting neutrinos from being affected by the flavour symmetry breaking in the composite sector. On the contrary, flavour symmetry breaking in the composite sector can be large  in Dirac models, 
provided that the tiny component of RH neutrinos in the elementary sector is flavour universal.

After a general presentation of the basic 4D flavourful HCHM, we pass to construct two explicit realizations in terms of 5D warped models. For concreteness, we consider the HCHM where the Higgs is a pseudo-Goldstone boson, i.e. gauge-Higgs unification models \cite{early}. 
The 5D models are based on the minimal $SO(5)\times U(1)_X$ gauge symmetry \cite{Agashe:2004rs}, while the flavour symmetry group contains, in addition to the $S_4\times \Z_3$ factor,  model-dependent discrete abelian factors necessary to minimize the number of allowed (and often unwanted) terms. 

In the Majorana model, the leading source of flavour violation arises from so called fermion boundary kinetic terms (BKT),
whose effect is analyzed in detail. The only sizable constraints come from lepton mixing, being LFV processes for charged leptons below the current bounds. We also argue that CP violating effects, such as the Electric Dipole Moments (EDM) for charged leptons, are negligibly small.
Keeping the prediction of the solar mixing angle $\theta_{12}$ within the experimentally allowed $3\sigma$ range requires
flavour symmetry breaking in the composite sector  to be at most of ${\cal O}(3\%\div 4\%)$ for neutrinos, unless a $\Z_2$ exchange symmetry is present on the IR brane, in which case no constraint occurs. This $\Z_2$-invariant 5D model is
surprisingly successful, simple and constrained, and essentially contains only one free real parameter and two Majorana phases! 
The model is compatible with the mass of the first Kaluza-Klein (KK) gauge resonances being $m_{KK}\gtrsim 3.5$ TeV, which is roughly the lowest scale allowed by electroweak considerations ($S$ parameter). The masses of all fermion KK resonances (charged and neutral) are always above the TeV scale.

In the Dirac model the most significant constraint does not arise from LFV processes or lepton mixing, but from
a too large deviation of the gauge coupling of neutrinos to the $Z$ boson from its SM value, which is constrained by LEP I to be roughly at the per mille level. This bound is satisfied by taking $m_{KK}\gtrsim $10 TeV, well above the LHC reach, with an ${\cal O}(1\%)$ tuning in the electroweak sector. 
The masses of charged fermion KK resonances  are above the TeV scale, while in the neutral fermion sector potentially light (sub-TeV) states can appear.

The structure of the paper is as follows. In section 2 we describe our set-up from a general effective
4D point of view both for the Majorana and Dirac models. In section 3 we briefly review the 
relevant operators entering in the LFV processes we focus on, radiative lepton decays $l_1\rightarrow l_2 \gamma$, decays to three leptons $l_1\rightarrow l_2 l_3 \bar l_4$ and $\mu-e$ conversion in nuclei.
In section 4 we construct the 5D Majorana model, compute its mass spectrum in subsection 4.1, 
the deviations from gauge coupling universality in subsection 4.2, LFV processes and lepton mixing in subsection 4.3 and estimate uncalculable effects in subsection 4.4.
In section 5 a similar, but more concise, analysis is repeated for the Dirac model.
We conclude in section 6. Three appendices are added. 
In appendix A basic definitions and properties of $S_4$ are reviewed, in appendix B we report our
conventions for the $SO(5)$ generators and representations and in appendix C we write the detailed structure of the two form factors governing the charged lepton radiative decays. 

%%%%%%%%%%%%%%%%%%%%%%%%%%%%%%%%%%%%%%%%%%%%

\section{General Set-up}
\label{sec:scenario}
%%%%%%%%%%%%%%%%%%%%%%%%%%%%%%%%%%%%%%%%%%%%

We consider CHM with a  non-abelian discrete flavour symmetry $G_f=S_4\times \Z_3$. They
consist of an ``elementary" and a ``composite" sector: 
\be
{\cal L}_{tot} = {\cal L}_{el} + {\cal L}_{comp}+{\cal L}_{mix}\,.
\ee
The symmetry $G_f$ is broken in the elementary sector  to $\Z_2\times \Z_2\times \Z_3$, where $\Z_2\times \Z_2 \subset S_4$ is generated by $S$ and $U$,  and  in the composite sector to $\Z_3^{(D)}$, the diagonal subgroup of the external $\Z_3$ and $\Z_3\subset S_4$ generated by $T$  (see appendix \ref{app:S4} for our notation and details on $S_4$ group theory).
We do not need to specify how the flavour symmetry breaking pattern is achieved.  
The term ${\cal L}_{mix}$ governs the mixing between the two sectors and is assumed to be invariant under the whole flavour group $G_f$.
This is our definition of HCHM in the following. We have two different classes of models, depending on whether
neutrino masses are of  Majorana or Dirac type. We will refer to the two cases as Majorana/Dirac  models (or scenarios).

\subsection{Majorana Models}

\label{subsec:Mscenario}

The elementary sector is invariant under the SM gauge group and 
 includes three generations of SM left-handed (LH) and RH leptons $l_L^\alpha$, $l_R^\alpha$ and three RH neutrinos $\nu_R^\alpha$. Here and in the following  Greek letters from the beginning of the alphabet denote generation indices; depending on the context, $\alpha=e,\mu,\tau$ or equivalently $\alpha =1,2,3$. The LH leptons $l_L^\alpha$ and the RH neutrinos $\nu_R^\alpha$ transform as $({\bf 3},1)$ under $S_4\times \Z_3$, while the RH leptons $l_R^\alpha$ transform as
$({\bf 1}, \omega^{2(\alpha-1)})$, where $\omega \equiv\rm{e}^{2\pi i/3}$ is the third root of unity. 
The elementary Lagrangian (up to dimension four terms) is taken to be 
\be
{\cal L}_{el} = \bar l_L^\alpha i \hat D l_L^\alpha + \bar l_R^\alpha i \hat D l_R^\alpha + \bar \nu_R^\alpha i \hat\partial \nu_R^\alpha - \frac 12  (\ol{\nu^{c}_R}^\alpha M_{\alpha\beta} \nu_R^\beta +h.c.)\,,
\label{Lele}
\ee
where the superscript $c$ denotes charge conjugation and $M$ is the most general mass matrix invariant under $\Z_2\times \Z_2 \times \Z_3$.  In flavour space, it is of the form 
\be
M = U_{TB} M_D U_{TB}^t \,,
\label{MUTB}
\ee
with $U_{TB}$ the TB mixing matrix
\begin{equation}
U_{TB} = \left( \begin{array}{ccc}
 \sqrt{\frac{2}{3}} & \sqrt{\frac{1}{3}} & 0\\
 -\sqrt{\frac{1}{6}} & \sqrt{\frac{1}{3}} & \sqrt{\frac{1}{2}}\\
 -\sqrt{\frac{1}{6}} & \sqrt{\frac{1}{3}} & -\sqrt{\frac{1}{2}}
\end{array}
\right)
\label{UTB}
\end{equation}
and $M_D$ a diagonal matrix. We use the notation $\hat A\equiv \gamma^\mu A_\mu$, for any vector $A_\mu$. 

The composite sector is an unspecified strongly coupled theory, that gives rise, among other states, to a composite
SM Higgs field. The latter may or may not be Goldstone fields coming from a spontaneously broken global symmetry.
 In absence of any interaction between the elementary and the composite sector, 
 the SM fermions are massless. They gain masses, after ElectroWeak Symmetry Breaking (EWSB), by mixing with
fermion operators $\Psi$ belonging to the strongly coupled sector. 
The mixing Lagrangian ${\cal L}_{mix}$ is
\be
{\cal L}_{mix}  =  \frac{\lambda_{l_L}}{\Lambda^{\gamma_{lL}}} \bar{l}_L^\alpha \Psi_{l_L,R}^\alpha + \frac{\lambda_{l_R}^\alpha}{\Lambda^{\gamma_{lR}^\alpha}} \bar{ l}_R^\alpha \Psi_{l_R,L}^\alpha +\frac{\lambda_{\nu_R}}{\Lambda^{\gamma_{\nu R}}} \bar{\nu}_R^\alpha\Psi_{\nu_R,L}^\alpha+h.c. 
\label{Lmix}
\ee
where $\Lambda$ is a high UV cut-off scale of the composite sector, $\Psi_{l_L}^\alpha$, $\Psi_{l_R}^\alpha$ and $\Psi_{\nu_R}^\alpha$ are fermion operators of (quantum) dimensions $5/2+\gamma_{lL}$, $5/2+\gamma_{lR}^\alpha$, $5/2+\gamma_{\nu R}$,  transforming as $({\bf 3},1)$, $({\bf 1}, \omega^{2(\alpha-1)})$ and $({\bf 3},1)$
under $S_4\times \Z_3$, respectively.
The mixing parameters $\lambda_{l_L}$ and $\lambda_{\nu_R}$ 
are flavour universal, while $\lambda_{l_R}^\alpha$ are flavour diagonal, but non-universal. For simplicity, 
we assume that all of them are real.
Although strictly not necessary, we take $\gamma_{lL}, \gamma_{lR}^\alpha>0$, so that
these mixing couplings are irrelevant. To a good approximation, $l_L^\alpha$ and $l_R^\alpha$
can be identified with the SM fields, with a small mixing with the strongly coupled sector.
Integrating out the composite fermion operators and taking into account that ${\cal L}_{comp}$ is invariant under $\Z_3^{(D)}$ only,  gives the following charged lepton mass matrix (in left-right convention, $\bar\psi_L M \psi_R$) 
\be
M_{l,\alpha\beta} \simeq \frac{\lambda_{l_L}}{\Lambda^{\gamma_{lL}}}\frac{\lambda_{l_R}^\beta}{\Lambda^{\gamma_{lR}^\beta}} \langle \bar\Psi_{l_R}^\beta \Psi_{l_L}^\alpha \rangle \sim 
   b_\alpha v_H\lambda_{l_L}\lambda_{l_R}^\alpha\delta_{\alpha\beta}\Big(\frac{\mu}{\Lambda}\Big)^{\gamma_{lR}^\alpha+\gamma_{lL}}\,,  \label{mlept4D}  
\ee
where $v_H$ is the electroweak scale, $\mu$ is the ${\cal O}$(TeV) scale at which the composite theory becomes strongly coupled and $b_\alpha$ are ${\cal O}(1)$ coefficients.\footnote{The estimate (\ref{mlept4D}) and the following are only valid for spontaneously broken CFT, which is the case mostly relevant for us. At a more qualitative level, however, our arguments apply to more generic holographic composite sectors, such as the ones in \cite{Panico:2010is}.}
The hierarchy of the charged lepton masses naturally arises from the $(\mu/\Lambda)$ suppression factor in (\ref{mlept4D}) with a proper choice of anomalous dimensions $\gamma_{lL}$ and $\gamma_{lR}^\alpha$.
The coupling $\lambda_{\nu_R}$ is in general relevant and $\nu_R^\alpha$ strongly mix
with the composite sector. The latter gives the following contribution to the kinetic terms of $\nu_R^\alpha$:
\be
\frac{\lambda_{\nu_R}^2}{\Lambda^{2\gamma_{\nu R}}} \!\!\int \!d^4 pd^4 q \, \bar \nu_R^\alpha(-p) \langle \Psi_{\nu_R,L}^\alpha(p) \bar\Psi_{\nu_R,L}^\beta(-q) \rangle \nu_R^\beta(q)  \sim \delta_{\alpha,\beta}\,\tilde b_\alpha^2
\lambda_{\nu_R}^2\Big(\frac{\mu}{\Lambda}\Big)^{2\gamma_{\nu R}}\!\!\!\int \! d^4 x \, \bar\nu_R^\alpha(x) i\hspace{0.05cm}  \hat\partial\, \nu_R^\alpha(x) \,,
\label{nuKinComp}
\ee
with $\tilde b_\alpha$ ${\cal O}(1)$ coefficients.  When $\gamma_{\nu_R}<0$, the kinetic term in (\ref{nuKinComp}) dominates over the ${\cal O}(1)$ term (\ref{Lele}) present in the elementary Lagrangian, and it is more appropriate to say that $\nu_R^\alpha$ are states in the composite sector with a small component in the elementary sector. 
When $\nu_R^\alpha$ are canonically normalized, the relevant coupling $\lambda_{\nu_R}$ in (\ref{Lmix}) becomes effectively a constant.\footnote{This can also be seen by solving an equation for the renormalization group flow of the couplings $\lambda$ \cite{Contino:2004vy,Agashe:2004rs}.}
The canonically normalized neutrino Dirac mass terms are of the form
\be
M_{\nu,\alpha\beta}^D \simeq \frac{\lambda_{l_L}}{\Lambda^{\gamma_{l L}}}\frac{\lambda_{\nu_R}}{\Lambda^{\gamma_{\nu R}}} 
\Big(\frac{\mu}{\Lambda}\Big)^{-\gamma_{\nu R}}\frac{1}{\tilde b_\beta \lambda_{\nu_R}}
\langle \bar\Psi_{\nu_R}^\beta \Psi_{l_L}^\alpha \rangle \sim 
\frac{\hat b_\alpha v_H\lambda_{l_L}}{\tilde b_\alpha} \delta_{\alpha\beta}\Big(\frac{\mu}{\Lambda}\Big)^{\gamma_{lL}}\,,  
\label{mneuDirac}
\ee
with $\hat b_\alpha$ ${\cal O}(1)$ coefficients.
Notice the crucial difference between the charged lepton (\ref{mlept4D}) and neutrino (\ref{mneuDirac}) masses.
The former explicitly break the flavour symmetry, since $\bar l^d_L l_R$ is not $S_4\times \Z_3$ invariant, while the latter do not, being $\bar l^u_L \nu_R$ an invariant. 
This implies that the coefficients $b_\alpha$ vanish in the limit of exact $S_4\times \Z_3$ symmetry, while $\hat b_\alpha = \hat b$,  $\tilde b_\alpha = \tilde b$ 
become flavour independent.
Assuming a small breaking of the flavour symmetry  
in the neutrino sector,
one can take  $\tilde b_\alpha\approx \tilde b$, $\hat b_\alpha\approx \hat b$ and, independently of $b_\alpha$, the Dirac neutrino mass terms (\ref{mneuDirac}) become universal. 
We stress the importance of having a small breaking of the flavour symmetry in the neutrino sector but not necessarily for charged leptons, because
the masses of the latter are already suppressed by their small degree of compositeness. Demanding a higher degree of compositeness, in particular for the $\tau$ lepton,  might result in too large deviations of its coupling to the $Z$ from its SM value.
Integrating out $\nu^\alpha_R$ in this limit gives the following see-saw like neutrino mass matrix:
\be
M_{\nu,\alpha\beta} \simeq \hat b^2 v_H^2 \lambda_{l_L}^2 \lambda_{\nu_R}^2 \Big(\frac{\mu}{\Lambda}\Big)^{2(\gamma_{\nu R}+\gamma_{l L})}
 \Big( U_{TB} M_D^{-1} U_{TB}^t\Big)_{\alpha\beta} \,,
\label{mMajo}
\ee
where we have again taken into account the scaling required to canonically normalize $\nu^\alpha_R$. Thanks to the latter, the factors $\tilde b$ cancel
from the final formula (\ref{mMajo}) but, more importantly, we gain a crucial enhancement factor $(\mu/\Lambda)^{2\gamma_{\nu_R}}$, without which  the light neutrino masses would be far too small for $M_D\sim {\cal O}(M_{Pl})$, $M_{Pl}$ being the reduced Planck mass, considering that the Dirac mass terms are at most of ${\cal O}(v_H)$. 
No intermediate mass scale has then to be advocated for $M_D$.

The mass matrix (\ref{mlept4D}) is diagonal in flavour space and no rotation of charged leptons is needed to go to the mass basis.
On the other hand,  the neutrino mass matrix (\ref{mMajo}) is diagonalized by the matrix $U_{TB}$ (up to phases), which leads to the identification
\be
U_{PMNS} = U_{TB}.
\ee

\subsection{Dirac Models}

\label{subsec:Dscenario}

The elementary sector includes, like in the Majorana scenario,  three generations of LH and RH leptons $l_L^\alpha$, $l_R^\alpha$ in $({\bf 3},1)$ 
and $({\bf 1}, \omega^{2(\alpha-1)})$ of $S_4\times \Z_3$, respectively, and in addition we now have LH exotic neutrino singlets $\hat \nu_L^\alpha$ in $({\bf 3},1)$.
The composite sector is assumed to contain two massless RH fermion bound states, singlets under $G_{SM}$, both  in $({\bf 3},1)$ of $S_4\times \Z_3$.
One of them mixes with $\hat\nu_L^\alpha$ giving rise to vector-like massive neutrinos $\hat \nu^\alpha$.  The remaining fermions, denoted by $\nu_R^\alpha$,
mix with some heavy vector-like states in the elementary sector. When the latter are integrated out, one is left with a tiny mixing mass term $\epsilon$ between  
$\nu_R^\alpha$ and $\hat \nu_L^\alpha$  in the elementary sector, which is flavour universal.\footnote{We do not specify here how this mixing is achieved and how its flavour universality is guaranteed. We will see that the latter requirement can naturally be fulfilled in our 5D example, see subsection 5.1. } The elementary Lagrangian (up to dimension four terms) is taken to be 
\be
{\cal L}_{el} = \bar l_L^\alpha i \hat D l_L^\alpha + \bar l_R^\alpha i \hat D l_R^\alpha + \bar{\hat \nu}_L^\alpha i\hat\partial \hat\nu_L^\alpha - \epsilon   (\bar{\hat\nu}^{\alpha}_L M_{\alpha\beta} \nu_R^\beta +h.c.)\,,
\label{LeleD}
\ee
with $M$ as in (\ref{MUTB}).
The mixing Lagrangian ${\cal L}_{mix}$ is 
\be
{\cal L}_{mix}  =  \frac{\lambda_{l_L}}{\Lambda^{\gamma_{lL}}} \bar{l}_L^\alpha \Psi_{l_L,R}^\alpha + \frac{\lambda_{l_R}^\alpha}{\Lambda^{\gamma_{lR}^\alpha}} \bar{ l}_R^\alpha \Psi_{l_R,L}^\alpha +\frac{\lambda_{\hat\nu_L}}{\Lambda^{\gamma_{\hat\nu L}}} \bar{\hat \nu}_L^\alpha\Psi_{\hat \nu_L,R}^\alpha+h.c. 
\label{LmixD}
\ee
The operators  $\Psi_{l_L}^\alpha$, $\Psi_{l_R}^\alpha$ and $\Psi_{\hat \nu_L}^\alpha$ are of dimensions $5/2+\gamma_{lL}$, $5/2+\gamma_{lR}^\alpha$, $5/2+\gamma_{\hat\nu L}$,  transforming as $({\bf 3},1)$, $({\bf 1}, \omega^{2(\alpha-1)})$ and $({\bf 3},1)$
under $S_4\times \Z_3$, respectively.
The mixing parameters $\lambda_{l_L}$ and $\lambda_{\hat\nu_L}$ are flavour universal, while $\lambda_{l_R}^\alpha$ are flavour diagonal, but non-universal. The charged lepton mass matrix 
is the same as (\ref{mlept4D}). The operators $\Psi_{\hat\nu_L,R}^\alpha$ excite, among other states, the RH massless neutrino bound states that pair up with $\hat \nu^\alpha_L$.
The vector-like mass of $\hat \nu^\alpha$ depends on the nature of the coupling $\lambda_{\hat \nu_L}$:
\bea
m_{\hat \nu}^\alpha & \sim & d_\alpha \lambda_{\hat\nu_L} \mu  \Big(\frac{\mu}{\Lambda}\Big)^{\gamma_{\hat \nu L}}, \ \ \ \ {\rm for } \; \gamma_{\hat \nu L} >0 \,, \nn \\
m_{\hat \nu}^\alpha & \sim & d_\alpha \lambda_{\hat\nu_L}  \mu \,, \hspace{1.8cm} {\rm for } \; \gamma_{\hat \nu L} <0\,,
\label{masshatnu}
\eea
where $d_\alpha$ are ${\cal O}(1)$ coefficients. When EWSB occurs, Yukawa couplings between $\Psi_{l_L}^\alpha$ and $\Psi^\alpha_{\hat \nu_L}$
induce mixing among $\nu_L^\alpha$ and $\hat \nu^\alpha$. When $\hat\nu^\alpha$ are integrated out, one gets
\be
\hat\nu_L^\alpha \sim \frac{\hat d_\alpha \lambda_{l_L} v_H}{m_{\hat\nu}^\alpha}  \Big(\frac{\mu}{\Lambda}\Big)^{\gamma_{l L}}
\nu_L^\alpha\,,
\label{nuhatINT}
\ee
where $\hat d_\alpha$ are  ${\cal O}(1)$ coefficients. 
Plugging (\ref{nuhatINT}) into the mass term in (\ref{LeleD}) gives the SM neutrino mass matrix
\be
M_{\nu,\alpha\beta}  \sim  \epsilon \frac{\hat d_\alpha \lambda_{l_L} v_H}{m_{\hat\nu}^\alpha}  \Big(\frac{\mu}{\Lambda}\Big)^{\gamma_{l L}}M_{\alpha\beta}\,.
\label{massDirac4d}
\ee
In the limit in which the masses $m_{\hat\nu}^\alpha$ and the mixing are universal, $m_{\hat\nu}^\alpha=m_{\hat\nu}$, $d_\alpha =d$, $\hat d_\alpha=\hat d$, the mass matrix (\ref{massDirac4d}) leads to TB mixing.
The composite nature of the RH neutrino naturally explains the smallness of $\epsilon$ and hence the actual SM neutrino masses \cite{ArkaniHamed:1999za}.
  
  \vskip 10pt 
In both scenarios, the  flavour symmetry $\Z_3^{(D)}$,  present in the composite sector, remains unbroken in the limit in which the neutrino mass term $M$ in the elementary sector is proportional to the identity. Correspondingly,  all tree-level flavour changing charged gauge boson interactions are vanishing in this limit. 
When $M$ is not proportional to the identity, the latter are still negligible  in the Majorana scenario, 
being suppressed by the masses of the heavy RH neutrinos, but can be sizable in the Dirac one, leading to processes such as $\mu\to e\gamma$. 
Tree-level flavour violating Higgs and neutral gauge boson interactions vanish  in both scenarios.
This summarizes the basic set-up of our Majorana and Dirac HCHM.
There are of course several sub-leading effects that should consistently be analyzed. We have not performed such analysis, but have preferred to postpone 
their discussion to the explicit 5D models that will follow. We only comment here that a relevant
source of flavour violation arises from the elementary sector, since the kinetic terms of the SM fermions (in the basis where ${\cal L}_{mix}$ is $G_f$ invariant)
are constrained in general to be only $\Z_2\times \Z_2\times \Z_3$ invariant, rather than $S_4 \times \Z_3$ invariant:
\be
 \bar l_L i \hat D l_L \rightarrow   \bar l_L (1+Z_l)  i \hat D l_L 
  \label{Z4D}
\ee
with $Z_l = U_{TB} Z_l^D U_{TB}^t$,  and $Z_l^D$ a diagonal matrix. 
Similar considerations apply of course to $\nu^\alpha_R$ and $\hat \nu^\alpha_L$, while  the additional unbroken $\Z_3$ symmetry forbids flavour violating kinetic terms for $l^\alpha_R$. 
As we will see, in 5D models the $Z_l$ factors are mapped to BKT at the UV brane.

%%%%%%%%%%%%%%%%%%%%%%%%%%%%%%%%

\section{Effective Field Theory for LFV Processes}
\label{sec:LFV}
%%%%%%%%%%%%%%%%%%%%%%%%

In this section we review, closely following \cite{Kuno:1999jp} and their notation, the most relevant effective operators entering in LFV processes.
The most experimentally constrained LFV observables are the radiative lepton decays $l_1\rightarrow l_2 \gamma$, the
decays to three leptons $l_1\rightarrow l_2 l_3 \bar l_4$ and the $\mu-e$ conversion in nuclei. Particularly relevant are the muon decays $\mu\rightarrow e\gamma $
and $\mu\rightarrow ee\bar e$ ($\mu\rightarrow 3 e$ for short). These LFV processes are described by the following
effective dimension 5 and 6 operators
\bea
-\frac{\sqrt{2}}{4G_F}{\cal L}_{eff} &\supset& m_\mu A_R \bar \mu_R \sigma^{\mu\nu} e_L F_{\mu \nu} + m_\mu A_L \bar \mu_L \sigma^{\mu\nu} e_R F_{\mu \nu} +
g_1 (\bar \mu_R e_L)(\bar e_R e_L) + g_2 (\bar \mu_L e_R)(\bar e_L e_R) \nn \\
&+& g_3 (\bar \mu_R \gamma^\mu e_R) (\bar e_R \gamma_\mu e_R)+ g_4 (\bar \mu_L \gamma^\mu e_L) (\bar e_L \gamma_\mu e_L)\nn \\
&+& g_5 (\bar \mu_R \gamma^\mu e_R) (\bar e_L \gamma_\mu e_L)+ g_6 (\bar \mu_L \gamma^\mu e_L) (\bar e_R \gamma_\mu e_R) + h.c. \,.
\label{LeffFCNC}
\eea
Terms of the form $(\bar \mu_R e_L)(\bar e_L e_R)$ and  $(\bar\mu_L  e_R )(\bar e_R e_L)$, by a Fierz identity, are shown to
contribute to $g_5$ and $g_6$, respectively. The first two terms contribute to $\mu\rightarrow e\gamma $ while all terms contribute  to $\mu\rightarrow 3e$.
One finds the following branching ratio for these processes:
\bea
BR(\mu\rightarrow e\gamma) & = & 384 \pi^2 (|A_L|^2+|A_R|^2)\,, \nn \\
BR(\mu\rightarrow 3e) &= & \frac{|g_1|^2+|g_2|^2}{8}+2(|g_3|^2+|g_4|^2)+|g_5|^2+|g_6|^2 + 
\label{BReff} \\
&& 8e {\rm Re}\big[ A_R (2g_4^*+g_6^*)+A_L (2g_3^*+g_5^*)\big]+64e^2 \Big(\log \frac{m_\mu}{m_e} - \frac{11}8\Big) (|A_L|^2+|A_R|^2)\,. \nn
\eea
The $\mu-e$ conversion in nuclei is more involved and described by an additional set of effective operators, that contain quark fields.
The most relevant ones are the vector 4 fermion operators:
\be
-\frac{\sqrt{2}}{G_F}{\cal L}_{eff} \supset \sum_{q=u,d} \Big[
 (g_{LV(q)}\bar e_L \gamma^\mu \mu_L +g_{RV(q)}\bar e_R \gamma^\mu \mu_R )\bar q \gamma_\mu q +h.c. \Big]\,.
 \label{Lagquark}
\ee
The branching ratio is given by  (see \cite{Kuno:1999jp} for more details)
\bea
B_{conv}(\mu N\rightarrow eN)\simeq &&\!\!\! \frac{m_\mu^5 G_F^2 F_p^2\alpha^3 Z_{eff}^4}{8\pi^2 Z\Gamma_{capt}}
\Big(|(2Z+N)g_{LV(u)}+(Z+2N)g_{LV(d)}|^2\nn \\
&+&  |(2Z+N)g_{RV(u)}+(Z+2N)g_{RV(d)}|^2\Big)\,,
\eea
where  $Z$ and $N$ are the proton and neutron numbers of the nucleus, $F_p$ is the nuclear form factor, $Z_{eff}$ is the effective atomic charge and $\Gamma_{capt}$ is the total muon capture rate.
A similar analysis applies to LFV processes involving the $\tau$ lepton, see e.g. \cite{Chang:2005ag} for details.

%%%%%%%%%%%%%%%%%%%%%%%%%%%%%%%%%%%%%%%%%%%%
\section{Explicit 5D Majorana Model}
\label{sec:leptonM}
%%%%%%%%%%%%%%%%%%%%%%%%%%%%%%%%%%%%%%%%%%%%

It is useful to construct a specific 5D weakly coupled description of our Majorana scenario, in order to concretely address its phenomenological viability beyond possible estimates based on na\"ive dimensional analysis only. We consider in the following a gauge-Higgs unification model in warped space \cite{Agashe:2004rs,Contino:2006qr}. As known, these models describe the sub-class of CHM where in  the composite sector (a spontaneously broken CFT) a global symmetry $G$ is spontaneously broken to a sub-group $H$, giving rise to a set of Goldstone fields including the SM Higgs field \cite{Contino:2003ve}.
We consider the minimal symmetry breaking pattern $SO(5)\rightarrow SO(4)$, leading only to the SM Higgs doublet. 
We use the conformally flat coordinates in which the 5D metric reads 
\be
\mathrm{d}s^2= a^2(z) \left( \eta_{\mu\nu} \mathrm{d}x^\mu \mathrm{d}x^\nu -\mathrm{d}z^2 \right) = 
\left( \frac{R}{z} \right)^2 \left( \eta_{\mu\nu} \mathrm{d}x^\mu \mathrm{d}x^\nu -\mathrm{d}z^2 \right)\,.
\ee
The UV and IR branes are located at $z=R\sim 1/M_{Pl}$, where $M_{Pl}$ is the reduced Planck mass,  and at $z=R^\prime \sim 1/$TeV, respectively.
The gauge symmetry in the bulk is 
\be
G_{\mathrm{gauge}}=SO(5)\times U(1)_X
\label{Gaugegroup}
\ee
and the flavour symmetry is
\be
G_{\mathrm{flavour}}=S_4\times \Z_3\times \Z^\prime_3\times \Z^{\prime\prime}_3 \,.
\ee
The gauge symmetry breaking is standard, with $G_{\mathrm{gauge}}$
broken at the UV and IR boundaries to  $G_{\mathrm{gauge, UV}}=SU(2)_L\times U(1)_Y$ and
$G_{\mathrm{gauge, IR}}=SO(4)\times U(1)_X\times P_{LR}$, where $P_{LR}$ is a LR $\Z_2$ symmetry, useful to
suppress deviations of the couplings of fermions to the $Z$ boson from their SM values \cite{Agashe:2006at}.
The flavour symmetry is broken
to $G_{\mathrm{flavour, UV}}=\Z_2\times \Z_2\times \Z_3\times \Z^{\prime\prime}_3$ and 
$G_{\mathrm{flavour, IR}}=\Z^{(D)}_3\times \Z^{\prime}_3$, 
respectively. 
In order to constrain  the number of terms allowed at the UV and IR boundaries,  
two additional symmetries $\Z^\prime_3$ and $\Z^{\prime\prime}_3$ have been included.

The lepton particle content of the model consists of 5D bulk fermions only: one fundamental $\xi_{l,\alpha}$, one adjoint
$\xi_{e,\alpha}$ and one singlet representation $\xi_{\nu, \alpha}$ of $SO(5)$, for each generation, all neutral under $U(1)_X$  (see \cite{Carena:2009yt} for a similar construction),
\begin{equation} \begin{array}{l}
\xi_{l,\alpha} = \left( \begin{array}{c}
\left[\, \wt{L}_{1,\alpha L} \, (-+) \, , \, L_{\alpha L} \, (++) \,\right]\\[0.2in] 
\hat{\nu}_{\alpha L} \, (-+)
\end{array}
\right), \ \  \\ \\
 \xi_{\nu,\alpha} = \nu_{\alpha L} \, (--) \end{array}
\xi_{e,\alpha} = \left( \begin{array}{c}
\begin{array}{c} x_{\alpha L} \, (+-)\\[0.05in] \wt{\nu}_{\alpha L} \, (+-)\\[0.05in] e_{\alpha L} \, (--)
\end{array}\,\,\,\,\,  Z_{\alpha L} \, (+-) \\[0.3in] 
\left[\, \wt{L}_{2,\alpha L} \, (+-) \, , \, \hat{L}_{\alpha L} \, (+-) \, \right]
\end{array}
\right),
\label{5Dfields}
\end{equation}
where the first and second entries in round brackets refer to the $+$ ($-$) Neumann (Dirichlet) boundary conditions (b.c.) at the UV and IR branes, respectively. 
We have written the $SO(5)$ multiplets in (\ref{5Dfields}) in terms of their $SU(2)_L\times SU(2)_R$ decomposition, where $[\psi_1,\psi_2]$
denotes the two components of the bi-doublet $({\bf 2},{\bf 2})$ with $T_{3R}=+1/2$ ($\psi_1$) and $T_{3R}=-1/2$ ($\psi_2$).
The SM LH lepton doublets arise from the zero modes of the 5D field $L_{\alpha L}$ in the ${\bf 5}$, the RH charged lepton singlets arise from the zero modes of $e_{\alpha R}$, $T_{3R}=-1$ component of the
$SU(2)_R$ triplet in the ${\bf 10}$, and the RH neutrinos arise from the zero modes of the singlet $\nu_{\alpha R}$.\footnote{The hypercharge $Y$ and electric charge $Q$ are given by $Y=X+T_{3R}$ and $Q=T_{3L}+Y$.}
 Notice that with the embedding (\ref{5Dfields}), the LH SM charged leptons, originating from 5D fields with $T_{3R} = T_{3L}= -1/2$, are expected to have suppressed SM $Z$ coupling deviations. 
In addition to the SM fields and their KK towers, the 5D fields (\ref{5Dfields}) give also rise to a set of exotic particles. In terms of $SU(2)_L\times U(1)_Y$, these are
two doublets $\wt{L}_{1,\alpha L}$ and $\wt{L}_{2,\alpha L}$ with $Y=1/2$, one doublet $\hat{L}_{\alpha L}$ with $Y=-1/2$, two singlets $\hat{\nu}_{\alpha L}$ and $\wt{\nu}_{\alpha L}$ with $Y=0$,
one singlet $x_{\alpha L}$ with $Y=1$ and one triplet $Z_{\alpha L}$ with $Y=0$.
The flavour properties of the fields (\ref{5Dfields}) are summarized in table \ref{table:flavourM}. Notice that the decomposition of the ${\bf 3}$ of $S_4$ into representations of the remnant group 
$\Z_2\times \Z_2$ at the UV boundary implies a non-trivial basis transformation, see appendix \ref{app:S4}.
\begin{table}
\begin{tabular}{|c|c|c|c|}\hline
& Bulk & UV & IR\\[0.05in]
\cline{2-4}
 & $G_{\mathrm{flavour}}=S_4\times \Z_3\times \Z_3^\prime\times \Z_3^{\prime\prime}$ & $G_{\mathrm{flavour,UV}}=\Z_2\times \Z_2\times \Z_3\times \Z_3^{\prime\prime}$ & \rule[0.1in]{0cm}{0cm} $G_{\mathrm{flavour,IR}}=\Z_3^{(D)}\times \Z_3^\prime$\\[0.05in]
\hline
 &   & \rule[0.05in]{0cm}{0cm} $(1,-1,1,\omega)$ & \\
 $\xi_{l,\alpha}$         &  $({\bf 3}, 1, \omega,\omega)$                                 & $(-1,1,1,\omega)$ & $(\omega^{2 (\alpha-1)},\omega)$\\
          &                                   & $(-1,-1,1,\omega)$ & \\[0.05in]
\hline
\rule[0.1in]{0cm}{0cm}$\xi_{e,\alpha}$ &  $({\bf 1}, \omega^{2 (\alpha-1)}, \omega, \omega)$ & $(1,1,\omega^{2 (\alpha-1)},\omega)$ & $(\omega^{2 (\alpha-1)},\omega)$\\[0.05in]
\hline
 &   & \rule[0.05in]{0cm}{0cm} $(1,-1,1,1)$ & \\
 $\xi_{\nu,\alpha}$         &  $({\bf 3}, 1, \omega,1)$                                 & $(-1,1,1,1)$ & $(\omega^{2 (\alpha-1)},\omega)$\\
          &                                   & $(-1,-1,1,1)$ & \\[0.05in]
\hline
\end{tabular}
\caption{Transformation properties of the 5D multiplets $\xi_l$, $\xi_e$ and $\xi_\nu$ under $G_{{\rm flavour}}$ and their decomposition properties under the subgroups $G_{\mathrm{flavour, UV}}$ and $G_{\mathrm{flavour, IR}}$ in the Majorana model.}
\label{table:flavourM}
\end{table}

The most general $G_{\mathrm{gauge, IR}} \times G_{\mathrm{flavour, IR}}$ invariant mass terms at the IR brane are\footnote{Following a common use in the literature, we have omitted to write certain fermion terms, including terms in which the would-be fermion components with Dirichlet b.c. appear, because
at the level of mass mixing the IR Lagrangian is only relevant in determining the modified b.c. (such as (\ref{IRbc}) below) of the fields at the IR brane. 
One can detect the absence of such terms by noticing that the variation of the sum of the bulk and brane action at the IR brane does not vanish when the b.c. (\ref{IRbc}) are imposed.}
\be
-\mathcal{L}_{\mathrm{IR}} = \left(\frac{R}{R'}\right)^4 \sum \limits_{\alpha=e,\mu,\tau}\left(m_{\mathrm{IR},\alpha}^l 
\left(\ol{\wt{L}}_{1,\alpha L} \wt{L}_{2,\alpha R} + \ol{L}_{\alpha L} \hat{L}_{\alpha R}\right)
+m_{\mathrm{IR},\alpha}^\nu  \, \ol{\hat{\nu}}_{\alpha L} \nu_{\alpha R} 
+ h.c. \right)\,,
\label{IRmasses}
\end{equation}
all flavour diagonal. The only $G_{\mathrm{gauge, UV}} \times G_{\mathrm{flavour, UV}}$ invariant mass terms at the UV brane 
are Majorana mass terms for RH neutrinos:
\begin{eqnarray}\nonumber
-\mathcal{L}_{\mathrm{UV}} &=& 
\frac{1}{12} \, m_{\mathrm{UV},e} \left( 2 \ol{\nu_{e R}^c} - \ol{\nu_{\mu R}^c} - \ol{\nu_{\tau R}^c} \right) 
\left( 2 \nu_{e R} - \nu_{\mu R} - \nu_{\tau R} \right)\\ \nonumber
&& \;\;\; + \frac{1}{6} \, m_{\mathrm{UV},\mu} \left( \ol{\nu_{e R}^c} + \ol{\nu_{\mu R}^c} + \ol{\nu_{\tau R}^c} \right) 
\left( \nu_{e R} + \nu_{\mu R} + \nu_{\tau R} \right)  
\\ \nonumber
&& \;\;\; + \frac{1}{4} \, m_{\mathrm{UV},\tau}  \left( \ol{\nu_{\mu R}^c} - \ol{\nu_{\tau R}^c} \right) 
\left( \nu_{\mu R} - \nu_{\tau R} \right) 
+ h.c.
\\ 
&=& \frac{1}{2} \, \ol{\nu_{\alpha R}^c} \mathcal{M}_{\mathrm{UV},\alpha \beta} \nu_{\beta R} + h.c.
\label{UVmasses}
\end{eqnarray}
with 
\be
\mathcal{M}_{\mathrm{UV}} = U_{TB}  m_{\mathrm{UV}} U_{TB}^t \,,
\label{MUV}
\ee
$m_{\mathrm{UV}}=\mbox{diag} \left( m_{\mathrm{UV},e}\, , \; m_{\mathrm{UV},\mu}\, , \; m_{\mathrm{UV},\tau} \right)$ and $U_{TB}$ as in (\ref{UTB}). Notice that the UV and IR localized mass terms are dimensionless.
The phases of the IR mass terms $m_{{\rm IR},\alpha}^l$ and $m_{{\rm IR},\alpha}^\nu$ can be removed by properly re-defining the 5D $SO(5)$ fields
$\xi_{l,\alpha}$ and $\xi_{e,\alpha}$. We can also remove one of the three phases of the UV mass terms $m_{{\rm UV},\alpha}$, so that, in total, the Majorana model contains just two phases.

\subsection{Mass Spectrum}

The mass spectrum of the theory (including all KK states) is efficiently computed using the so-called holographic approach
\cite{Luty:2003vm}, which is also very useful to match the 5D theory to the 4D description given in section 2.
As far as the lightest modes are concerned, however, simple and reliable formulas are more easily obtained using the more standard KK approach
and the so called Zero Mode Approximation (ZMA), which we use in the following. 
The ZMA is defined as the approximation in which EWSB effects (i.e. Higgs insertions) are taken as 
perturbations and mixing with the KK states coming from Higgs insertions is neglected.
The spectrum of the zero modes is then entirely fixed by the unperturbed zero mode wave functions and their overlap with the Higgs field. 
These unperturbed wave functions satisfy the new b.c. as given by the localized IR terms.
As explained in \cite{Huber:2003sf}, the localized UV Majorana mass terms, instead, must be considered as a perturbative mass insertion (like the Higgs) if one wants to
recover a meaningful mass spectrum for the light SM neutrinos without taking into account mixing with the KK states.
Due to the wave function localization of zero and KK modes, as a rule of thumb, the lighter the zero mode masses are, the more accurate the ZMA is. 

Taking into account the localized IR mass terms (\ref{IRmasses}), the IR b.c. for the non-vanishing 5D field components in ZMA are 
\begin{eqnarray}
\hat{\nu}_{\alpha R} &=& - m_{\mathrm{IR},\alpha}^\nu \, \nu_{\alpha R}, \ \ \ \ 
\nu_{\alpha L} =  m_{\mathrm{IR},\alpha}^{\nu} \, \hat{\nu}_{\alpha L} \,,   \ \ \ z= R^\prime \nn \\
 L_{\alpha R}& = & -m_{\mathrm{IR},\alpha}^l \hat L_{\alpha R} \,, \ \ \ \hat L_{\alpha L} = m_{\mathrm{IR},\alpha}^{l}  L_{\alpha L}\,, \ \ \ z= R^\prime\,.
 \label{IRbc}
\end{eqnarray}
We get the following zero mode expansion:
\begin{eqnarray}
L_{\alpha L}(x,z) &=& \frac{1}{\sqrt{R'}} \left( \frac{z}{R}\right)^2 \left( \frac{z}{R'}\right)^{-c_l}
f_{c_l} \frac{1}{\sqrt{\rho_\alpha}} l_{\alpha L}^{(0)} (x) \,, \label{LexpKK0}
\\ 
\hat{L}_{\alpha L}(x,z) &=& m_{\mathrm{IR},\alpha}^{l} \frac{1}{\sqrt{R'}} \left( \frac{z}{R}\right)^2 \left( \frac{z}{R'}\right)^{-c_\alpha}
f_{c_l} \frac{1}{\sqrt{\rho_\alpha}} l_{\alpha L}^{(0)} (x)\,,\label{LhatexpKK0}  \\
\nu_{\alpha R}(x,z)  &=& \frac{1}{\sqrt{R'}} \left( \frac{z}{R}\right)^2 \left( \frac{z}{R'}\right)^{c_\nu}
f_{-c_\nu} \frac{1}{\sqrt{\sigma_\alpha}} N_{\alpha R}^{(0)} (x) \,, \label{nuRexpKK0}
\\ 
\hat{\nu}_{\alpha R}(x,z)  &=& -m_{\mathrm{IR},\alpha}^\nu \frac{1}{\sqrt{R'}} \left( \frac{z}{R}\right)^2 \left( \frac{z}{R'}\right)^{c_l}
f_{-c_\nu} \frac{1}{\sqrt{\sigma_\alpha}} N_{\alpha R}^{(0)} (x)\,, \label{nuhatRexpKK0} \\
e_{\alpha R}(x,z)  & = & \frac{1}{\sqrt{R'}} \left( \frac{z}{R}\right)^2 \left( \frac{z}{R'}\right)^{c_\alpha} f_{-c_\alpha} e_{\alpha R}^{(0)} (x)\,, \label{eRexpKK0} 
\end{eqnarray}
where $l_{\alpha L}^{(0)}(x)$, $e_{\alpha R}^{(0)}(x)$ and $N_{\alpha R}^{(0)}(x)$ are the canonically normalized LH lepton doublets, RH charged leptons and RH neutrino zero modes, respectively.
We use the standard notation
\begin{equation}
f_{c} = \left[ \frac{1-2c}{1- \left( \frac{R}{R'}\right)^{1-2c}} \right]^{1/2}
\label{fcDEF}
\end{equation}
where $c = M R$ are the dimensionless bulk mass terms of the 5D fermions.  We denote by $c_l$ and $c_\nu$ 
the bulk mass terms of $\xi_{l,\alpha}$ and $\xi_{\nu,\alpha}$, constrained by the flavour symmetry to be flavour-independent.
We denote by $c_\alpha$ the remaining 3 bulk mass terms for $\xi_{e,\alpha}$.
The parameters $\rho_\alpha$ and $\sigma_\alpha$ are defined as
\begin{equation}
\rho_\alpha = 1 + |m_{\mathrm{IR},\alpha}^l|^2 \left( \frac{f_{c_l}}{f_{c_\alpha}} \right)^2\,, \ \ \ \ \ 
\sigma_\alpha = 1 + |m_{\mathrm{IR},\alpha}^\nu|^2 \left( \frac{f_{-c_\nu}}{f_{-c_l}} \right)^2\,.
\label{rhogammaDEF}
\end{equation}
We take the unitary gauge for the $SO(5)\rightarrow SO(4)$ symmetry breaking pattern in which the Higgs field wave function is (see appendix \ref{app:SO5} for our $SO(5)$ conventions) 
\be
A^{\hat a}_5(x,z) =\sqrt{\frac{2}{R}}\frac{z}{R^\prime} \langle h^{\hat a}(x) \rangle  = v_H \sqrt{\frac{2}{R}}\frac{z}{R^\prime}\delta ^{\hat a, 4}\ \equiv v_H f_H(z) \delta ^{\hat a, 4}\,,
\ee
with $v_H\simeq 250$ GeV. We find useful to introduce
\be
h \equiv \frac{v_H}{f_H}, 
\ee
where $f_H$ is the Higgs decay constant. It is defined as
\be
f_H = \frac{2\sqrt{R}}{g_5R^\prime} = \frac{2}{gR^\prime \sqrt{\log(R^\prime/R)}}\,.
\ee
In the second equality we have used the approximate tree-level matching between the $SO(5)$ 5D coupling $g_5$ and the $SU(2)_L$ 4D coupling $g$
\be
g_5 = g \sqrt{R \log (R'/R)}\,.
\ee
Flavour-independent bounds, essentially the $S$ parameter in models with a custodial symmetry, 
constrain $1/R^\prime\gtrsim 1.5$ TeV, corresponding to  $h \lesssim 1/3$.

By computing the wave-function overlap with the Higgs field, we get the following charged lepton mass matrix
\begin{equation}
M_{l, \alpha\beta} = \frac{h}{\sqrt{2}R^\prime}  \, f_{c_l} f_{-c_\alpha} \frac{m_{\mathrm{IR},\alpha}^l}{\sqrt{\rho_\alpha}} 
\delta_{\alpha,\beta}\,.
\label{CLMassMatrix}
\end{equation}
As usual, the SM fermion masses are naturally obtained by taking $c_\alpha<-1/2$, in which case $f_{-c_\alpha}$ are exponentially small
and hierarchical.

For the Dirac neutrino mass matrix we get
\begin{equation}
M_{\nu, \alpha\beta}^D = \frac{i h}{\sqrt{2}R^\prime}  \, f_{c_l} f_{-c_\nu} \frac{m_{\mathrm{IR},\alpha}^\nu}{\sqrt{\rho_\alpha \, \sigma_\alpha}} 
\delta_{\alpha,\beta} \,.
\label{MassaDirac5D}
\end{equation}
The Majorana mass matrix in 4D is the one on the UV boundary. Taking into account the wave functions of the 
RH neutrinos, we get
\begin{equation}
\mathcal{M}_{M,\alpha\beta} = \left( \frac{R}{R'}\right)^{2 c_\nu+1} f_{-c_\nu}^2 
\frac{1}{\sqrt{\sigma_\alpha}} \frac{\mathcal{M}_{\mathrm{UV}, \alpha\beta}}{R} \frac{1}{\sqrt{\sigma_\beta}} \,.
\end{equation}
Integrating out the heavy Majorana fields $N_{\alpha R}^{(0)}(x)$, the factors $\sigma_\alpha$ cancel out 
and the actual form of the light neutrino mass matrix is, using (\ref{MUV}),
\be
M_{\nu,\alpha\beta} =\frac{h^2}{2R^{\prime 2}} f_{c_l}^2 
\left(\frac{R'}{R} \right)^{2 c_\nu+1} \frac{m_{\mathrm{IR},\alpha}^\nu}{\sqrt{\rho_\alpha}} 
\Big(U_{TB}\frac{R}{m_{\mathrm{UV}}} U_{TB}^t\Big)_{\alpha\beta} \frac{m_{\mathrm{IR},\beta}^\nu}{\sqrt{\rho_\beta}}\,.
   \label{neutrinomassZMA}
\ee
For $m_{\mathrm{IR},\alpha}^\nu\approx m_{\mathrm{UV},\alpha}\approx {\cal O}(1)$, the size of the neutrino masses is mainly governed by the bulk mass term $c_\nu$. The latter is essentially fixed to be 
\be
c_\nu \approx - 0.36  \,.
\label{cnu}
\ee
We have explicitly checked that the masses of the zero modes obtained in the ZMA (and treating the UV Majorana mass term as a perturbative mass insertion)
are in excellent agreement with the exact tree-level spectrum. 

Let us consider the relation between the 5D model and the general 4D analysis performed in subsection \ref{subsec:Mscenario}. The strongly coupled sector is a CFT spontaneously broken at the scale $\mu\simeq 1/R^\prime$ with a cut-off $\Lambda \simeq 1/R$. The anomalous dimensions appearing in (\ref{Lmix})  are uniquely fixed by the bulk masses of the 5D multiplets $\xi_l$, $\xi_e$ and $\xi_\nu$ \cite{Contino:2004vy}:\footnote{The IR localized mass terms (\ref{IRmasses}) correspond to irrelevant deformations of the CFT and do not  affect the anomalous dimensions computed in the limit of exact conformal symmetry. They however deform the mass spectrum of the CFT when finite cut-off effects are taken into account.} 
\be
\gamma_{lL} = |c_l +1/2|-1, \ \ \ \gamma^\alpha_{l R} = |c_\alpha -1/2|-1, \ \ \  \gamma_{\nu R} = |c_\nu -1/2|-1.
\ee
It is straightforward to show that, for $c_l>1/2$ and $c_\alpha<-1/2$, the $(\mu/\Lambda)$ suppression factors appearing in (\ref{mlept4D}) arise from the factors $f_{c_l}$ and $f_{-c_\alpha}$ defined in (\ref{fcDEF}) and that $b_\alpha \sim m_{{\rm IR},\alpha}^l$.
With the value of $c_\nu$ taken as in (\ref{cnu}), the coupling $\lambda_{\nu_R}$ is relevant and $f_{-c_\nu}\simeq f_{c_\nu}\sim {\cal O}(1)$.
The mass formula (\ref{MassaDirac5D}) is of the general form (\ref{mneuDirac}), where $\hat b_\alpha \sim m_{\mathrm{IR},\alpha}^\nu$ and $\tilde b_\alpha^2 \sim \sigma_\alpha-1$.
 The latter factors, as expected, do not appear in the final mass formula (\ref{mMajo}).
The $\rho_\alpha$ are wave function normalization factors that take into account the contribution of the composite sector to the kinetic terms of the LH doublets, given by $\rho_\alpha-1$.

In the limit of an $S_4$ invariant IR Lagrangian, $m_{\mathrm{IR},\alpha}^l\rightarrow 0$ (so that $\rho_\alpha\rightarrow 1$) and $m_{\mathrm{IR},\alpha}^\nu=m_{\mathrm{IR}}^\nu$,
the neutrino mass matrix (\ref{neutrinomassZMA}) leads to TB mixing. As we will see, bounds on gauge coupling deviations 
favour the region in parameter space where $c_l$ is close to $1/2$, in which case $\rho_\alpha$ is equal to one (since $f_{c_\alpha}\gtrsim 1$) to a reasonable approximation even for $m_{\mathrm{IR},\alpha}^l\sim {\cal O}(1)$.
This accidental property allows us to also explore the region in parameter space where the flavour symmetry breaking in the charged lepton sector on the IR brane is large, while in the neutrino sector it remains small, namely $m_{\mathrm{IR},\alpha}^\nu=m_{\mathrm{IR}}^\nu(1+\delta m_{\mathrm{IR},\alpha}^\nu)$, with $\delta m_{\mathrm{IR},\alpha}^\nu\ll1$.

Instead of assuming a small breaking in the neutrino sector and for the sake of reducing the number of parameters in the model, one might also advocate an accidental
$\Z_2$ exchange symmetry present only in the IR localized Lagrangian, under which 
\be
\hat \nu_\alpha(x,R^\prime) \leftrightarrow \nu_\alpha(x,R^\prime) \,.
\label{Z2acc}
\ee
If the symmetry (\ref{Z2acc}) is imposed, the IR mass parameters $m_{\mathrm{IR},\alpha}^\nu$ are constrained to be equal to $\pm 1$.
Among the four inequivalent choices of $\pm 1$, we can take the universal choice $m_{\mathrm{IR},\alpha}^\nu=1$. 
Although not necessary, an analogous $\Z_2$ symmetry exchanging the two bi-doublets in the ${\bf 5}$ and the ${\bf 10}$ of $SO(5)$ (a single $\Z_2$
exchanging the bi-doublets and the singlets is also a viable possibility) might be advocated to also set $m_{\mathrm{IR},\alpha}^l=1$.
The resulting model can be seen as an ultra-minimal 5D model, with in total only 8 real parameters (5 bulk mass terms and 3 localized UV mass parameters) and two
phases (contained in the UV mass parameters), 4 of which
 are essentially fixed by the SM charged leptons ($c_\alpha$,  $c_l$), 1 by the overall neutrino mass scale ($c_\nu$) and 2 by the neutrino mass square differences
 (two combinations of $m_{\mathrm{UV},\alpha}$), leaving in this way just one free real parameter and two Majorana phases! We denote this constrained model by the ``$\Z_2$-invariant" model.

The mass spectrum of all the KK resonances is above the TeV scale. For instance, let us consider the $\Z_2$-invariant model and let us take $h=1/3$, $c_l=0.52$, $c_\nu=-0.365$ as benchmark values. In this case, 
the lightest gauge boson KK resonances have $m_{KK} \approx 3.5$ TeV, the lightest (negatively and positively) charged and neutral fermions have  masses around $2$ TeV, while  the heavy Majorana neutrinos have masses around $10^{13}$ GeV.

\subsection{Deviations from Gauge Coupling Universality}

In this subsection we compute the deviations from the SM values of the couplings of leptons to the $Z$ and $W$ bosons.  
In RS-like models such deviations can play an important role, since their expected order of magnitude for natural models
with $1/R^\prime\gtrsim 1.5$ TeV can be of the same order of magnitude or larger than the experimental bounds, which are at the per mille level. The size of the deviation is mainly fixed by the wave function profile in the fifth dimension of the 4D lepton.
The more the field is UV peaked, the smaller the deviation is. On general grounds, one might expect
sizable deviations for all the $Zl_L\bar l_L$ couplings and for the $Z\tau_R\bar\tau_R$ one.
Deviations of the LH neutrino couplings $Z\nu_L\bar \nu_L$ should also be studied. The latter have indirectly been measured by LEP I and are constrained at the per mille level with an accuracy comparable to that for charged leptons, using the invisible decay width of the $Z$ boson, under the assumption that this is entirely given by neutrinos \cite{LEPZnunu}.

An efficient way to compute these deviations, automatically summing over all the KK contributions, is provided 
by the holographic approach. In the latter, the effective 4D gauge couplings between fermions and gauge bosons are obtained by integrating over the internal dimension the 5D gauge vertex, with the 5D fields replaced by bulk-to-boundary
propagators and 4D fields. The main source of deviation arises from higher-order operators with Higgs insertions, which give
a contribution of ${\cal O}(h^2)$.  Higher-order derivative operators are negligible, being suppressed by the fermion masses or $Z$ boson mass and are ${\cal O}(M_l R')^2$ or ${\cal O}(m_Z R')^2$,  respectively.
The momentum of all external fields (and hence of all bulk-to-boundary propagators) can be then reliably set to zero. In this limit, the computation greatly simplifies and compact analytic formulae can be derived. In the following we do not report all the details of our computation but only the final results.
We define the 4D SM couplings $g_{l,SM}$ as 
\be
g_{l,SM} = T_{L}^3- Q \sin^2\theta_W\,,
\ee
without additional factors of the coupling $g$ or of the weak mixing angle $\theta_W$.

Let us first consider the LH charged leptons $l^\alpha_L$. Given our embedding of $l^\alpha_L$ into 5D multiplets with $T_{3L}=T_{3R}$, we simply have 
\be
\delta g_{l_L}^\alpha = g^\alpha_{l_L}-g^\alpha_{l_L,SM} = 0
\label{deltagL}
\ee
and no deviations occur at all.\footnote{The coupling deviations above are defined in the field basis in which a completely localized UV fermion has SM gauge couplings, with no deviations. In this basis the fermion independent universal coupling deviation arising from gauge field mixing is encoded in the $S$ parameter.}
They occur for the RH charged leptons $l^\alpha_R$. We get
\be
\delta g^{\alpha}_{l_R} \simeq - (M_{l, \alpha} R^\prime)^2 \, f_{c_l}^{-2} \frac{ (2+4 c_l+(3+2 c_\alpha) |m_{\mathrm{IR},\alpha}^l|^2)}{2 |m_{\mathrm{IR},\alpha}^l|^2(3+2c_\alpha)(1+2c_l)}\,,
\label{deltagR2}
\ee
where $M_{l,\alpha}$ are the charged lepton masses (\ref{CLMassMatrix}) and it is understood that $c_\alpha$ entering in (\ref{deltagR2}) are determined as a function of $c_l$, $m_{\mathrm{IR},\alpha}^l$ and $M_{l,\alpha}$. Equation (\ref{deltagR2}) clearly shows that a small flavour symmetry breaking in the composite sector for charged leptons, i.e. $m_{\mathrm{IR},\alpha}^l\ll 1$, is disfavoured. Keeping $c_l$ and $M_{l,\alpha}$ fixed, for small IR mass terms $\delta g^{\alpha}_{l_R}\propto 1/|m_{\mathrm{IR},\alpha}^l|^2$.
It is intuitively clear that $\delta g^{\alpha}_{l_R}$ grow when the localized IR mass terms decrease, since one needs to delocalize more the RH leptons to get their correct masses, resulting in larger mixing with the KK spectrum and hence larger deviations. Alternatively, one has to decrease the value of $c_l$, 
increasing the degree of compositeness of the LH leptons.

Let us now turn to the neutrino $Z$ couplings. 
Since $\nu^\alpha_L$ are embedded into $SO(5)$ multiplets with $T_{3L}\neq T_{3R}$, non-trivial deviations are expected.
In the limit $R^\prime\gg R$ and in the relevant range $c_l>0$, $c_\alpha <0, c_\nu<0$, we have
\be
\delta g_{\nu_L}^\alpha \simeq  \frac{h^2(1-2c_l)\Big( 4c_\alpha-2 + |m_{\mathrm{IR},\alpha}^l|^2 (2 c_l -3)+|m_{\mathrm{IR},\alpha}^\nu|^2\frac{(2c_\alpha-1) (2c_l-3)}{(2c_\nu-1)} \Big) R^\prime R^{2c_l}}{
4(2c_l-3 ) \Big(\big(2c_\alpha -1+ |m_{\mathrm{IR},\alpha}^l|^2 (2c_l-1)\big) R^\prime R^{2 c_l} + 
  R (R^\prime)^{2c_l}  (1 - 2 c_\alpha)\Big)}\,.
\label{deltagnu}
\ee
The couplings of the $W$ boson to the LH doublets and their deviations from the SM values, denoted by  $\delta g^\alpha_{\nu_L l_L}$, are computed in the same way. In the same limit as (\ref{deltagnu}), we find
\be
\left|\frac{\delta g^\alpha_{\nu_L l_L}}{g^\alpha_{\nu_L l_L}}\right| = \frac 12 \left|\frac{\delta g^\alpha_{\nu_L}}{g_{\nu_L}^\alpha}\right| \,.
\ee
We demand that 
\be
\left|\frac{\delta g^\alpha_l}{g^{\alpha}_l}\right|  < 2 \, \promille \,, \ \ \ \ \left|\frac{\delta g^{\alpha}_\nu}{g_\nu^\alpha}\right| < 4 \, \promille 
\label{EWPbounds}
\ee
for LH and RH charged leptons and LH neutrinos.  

The LH neutrino deviations (\ref{deltagnu}) are mostly sensitive to $c_l$, requiring  $c_l\gtrsim 0.49$, with a mild dependence on the other parameters, while the RH charged lepton deviations are also very sensitive to the bi-doublet IR mass parameters, disfavouring small values of $m_{\mathrm{IR},\alpha}^l$. Independently of $m_{\mathrm{IR},\alpha}^l$, we get an upper bound on $c_l$ from the $\tau$ lepton, $c_l\lesssim 0.56$. As in many warped models with bulk fermions, the region $c_l \simeq 1/2$ is preferred by electroweak bounds.

\subsection{LFV Processes and BKT}

\label{subsec:Fbounds}

Due to our choice of discrete symmetries, no 5D operators that reduce to the operators appearing in (\ref{LeffFCNC}) and in (\ref{Lagquark}) are allowed in the bulk or on the IR brane. The flavour preserving dipole operators responsible for lepton EDM are also forbidden by gauge invariance.
Operators associated with the couplings $g_4$ and $g_6$ in  (\ref{LeffFCNC}) and $g_{LV(q)}$ in (\ref{Lagquark}) are allowed on the UV brane, but their natural scale is ${\cal O}(M_{Pl}^{-2})$ and thus totally negligible. The operators in (\ref{LeffFCNC}) and in (\ref{Lagquark}) can only arise in the effective field theory below the KK scale, after the KK resonances have been integrated out. Their coefficients are then calculable.  In absence of further corrections, tree-level flavour changing interactions among charged leptons mediated by neutral KK gauge bosons and Higgs vanish, since all interactions and Yukawa couplings involving charged leptons are manifestly flavour diagonal (in contrast to what happens 
in generic RS models \cite{Agashe:2006iy}):
\be
g_{1-6}  = g_{LV(q)} = g_{RV(q)} = 0\,.
\ee 
Flavour violation occurs in the neutrino sector and hence radiative decays mediated by neutrinos and charged gauge bosons 
do not vanish, $A_L, A_R\neq 0$. However, these are negligible, because effectively mediated only by heavy Majorana neutrinos.
This is best seen by considering again the UV Majorana mass term as a mass insertion, but beyond ZMA, including all KK wave functions.
The mass terms (\ref{UVmasses}) can be written as follows:
\be
\label{eq:LMajinitial}
{\cal L}_{4D}^{Maj} = \frac 12  \Big(\sum _{m=0}^\infty  \ol{N_{\alpha R}^{(m)c}}(x)   f_{\nu,\alpha R}^{(m)}(R)  \Big)\mathcal{M}_{\mathrm{UV},\alpha\beta}  \Big(\sum _{n=0}^\infty   f_{\nu,\beta R}^{(n)}(R)  N_{\beta R}^{(n)}(x) \Big) +h.c.
\ee
where $f_{\nu,\alpha R}^{(n)}$ are the KK wave functions for the fields $N_{\alpha R}^{(n)}$, 
and explicitly show that the Majorana mass matrix has rank 1 in the KK indices, for each flavour $\alpha$.
The heavy Majorana state $N_{\alpha R}^h$ is defined by the eigenvector in round brackets in (\ref{eq:LMajinitial}), with non-vanishing Majorana mass. Suitable orthonormal combinations of $N_{\alpha R}^{(n)}$ define the ``light" KK modes $N_{\alpha R}^{l(n)}$.
The fields $N^h_{\alpha R}$ are not yet in their mass basis and (\ref{eq:LMajinitial}) is not diagonal in flavour space. However, since these
fields are very heavy, we can integrate them out. In the limit of infinite mass,  this implies setting $N_{\alpha R}^h=0$. Eventually, we see that 
the remaining terms in the Lagrangian involving the fields $N_{\alpha R}^{l(n)}$ are flavour-diagonal with real coefficients. 
For finite mass, flavour {\it and} CP violating interactions are generated, but suppressed by the heavy Majorana mass and
are completely negligible. 

It is important to study at this stage the impact of higher dimensional flavour violating operators in the model. These can only occur 
at the UV brane.  The lowest dimensional operators of this form are fermion BKT. 
In principle all possible BKT allowed by the symmetries must be considered. In practice this is rather difficult to do, so we focus only on those BKT,  whose presence with all others set to zero, causes flavour violation. From the table \ref{table:flavourM}, we see that $\Z_3$ forbids the appearance of flavour violating  BKT for $\xi_{e,\alpha}$, while these are allowed for $\xi_{l,\alpha}$ and $\xi_{\nu,\alpha}$.   
There are in principle four possible flavour violating BKT at the UV brane, for  $\tilde L_{1,\alpha R}$, $\hat \nu_{\alpha R}$, $\nu_{\alpha R}$  and $L_{\alpha L}$.
The KK expansion of fields with b.c. modified by both boundary mass and kinetic terms is quite involved. 
In order to simplify the analysis, we consider the BKT as a perturbation and treat them as insertions, like the Majorana mass terms. Namely, we take as b.c. for all fields the ones
with vanishing BKT and then plug the resulting KK expansion into the BKT. This approximation is clearly valid for parametrically small BKT, but it is actually very good at the
UV brane even for BKT of ${\cal O}(1)$, as we will see  (see \cite{Carena:2004zn} for an analysis of fermion BKT in warped models). Among the 4 BKT above, the UV BKT for $\tilde L_{1,\alpha R}$, $\hat \nu_{\alpha R}$ and $\nu_{\alpha R}$ are strongly suppressed (at least for the most relevant low KK modes), due to the form of the wave functions of these fields, and can be neglected. We are only left with
\be
{\cal L}_{BKT} = \bar L_L(x,R) (R \hat Z_l) i\hat D L_L(x,R) \,,
\label{LBKT}
\ee 
where $\Z_2\times \Z_2$ constrains $\hat Z_l$ to be of the form $\hat Z_l = U_{TB} {\rm diag} \,(z_{el},z_{\mu l},z_{\tau l}) U_{TB}^t$. The coefficients $z_{\alpha l}$ are dimensionless and their natural values are ${\cal O}(1)$, although smaller values $\sim 1/(16\pi^2)$ can also be radiatively stable. If one assumes a small breaking of $S_4\rightarrow \Z_2\times \Z_2$ at the UV brane,
the relative differences in the $z_{\alpha l}$ can be taken parametrically smaller than $\sim 1/(16\pi^2)$. 
In presence of the flavour violating operators (\ref{LBKT}), the couplings (\ref{LeffFCNC}) and (\ref{Lagquark}) become non-vanishing. 

Let us first write down, in the mass basis,  the relevant interaction terms of our 5D Lagrangian
that give rise to the  effective couplings present in  (\ref{LeffFCNC}) and (\ref{Lagquark}). We have
\bea
{\cal L}_{int}& \supset &\frac{g}{\sqrt{2}}\Bigg[\sum_{i,a,V^-}\!\! \Big(C_{iL}^{a} \bar l_{a L} \hat V^-  \nu_{iL} + C_{iR}^{a} \bar l_{a R} \hat V^- \nu_{iR} + h.c. \Big)+\sum_{q,V^0} g_{q} \bar q \hat V^0 q   \label{IntLagLFV}   \\
&+& \sum_{a,b,V^0}\!\! \Big( D_L^{a b} \bar l_{a L} \hat V^0 l_{ b L} + D_R^{a b} \bar l_{a R} \hat V^0 l_{ b R} \Big) +\sum_{a, b}
   \Big(Y_{a b} \bar l_{a L} H l_{ b R} + Y_{ ba}^* \bar l_{a R} H l_{ b L}  \Big)\Bigg] , \nn
\eea
where $a$ and $ b$  run over all charged leptons, $q$ runs over the light SM quarks $u$ and $d$, $i$ runs over all the neutrinos, $V^-$ and $V^0$ run over all charged and neutral gauge fields, respectively. By ``all" we here mean all species of particles, including their KK resonances. For simplicity of notation, we have omitted the implicit dependence of the couplings in (\ref{IntLagLFV}) on the gauge fields $V^-$ and $V^0$.
The couplings in (\ref{Lagquark}) depend on how the quark sector is realized in the theory.
We assume here that up and down quarks are genuine 4D fields localized at the UV brane and singlets under the flavour symmetry.

The coefficients $A_L$ and $A_R$ are radiatively generated and receive contributions from  3 different classes of one-loop diagrams, 
\be
A_R = A_R^{(W)} + A_R^{(Z)} + A_R^{(H)}\,, \ \ \ \ \ A_L = A_L^{(W)} + A_L^{(Z)} + A_L^{(H)}\,, 
\label{ALAReff}
\ee
where $A_{R/L}^{(W)}$ are the contributions due to the diagrams where a charged gauge boson
and a neutrino are exchanged, $A_{R/L}^{(Z)}$ are the contributions due to the diagrams where a neutral gauge boson
and a charged lepton are exchanged and $A_{R/L}^{(H)}$ are the contributions due to the diagrams where the Higgs
and a charged lepton are exchanged. The explicit form of these coefficients is reported in appendix \ref{app:ALAR}.

The operators associated with the couplings $g_{1-6}$ are generated at tree-level by Higgs and neutral gauge boson exchange. By matching, we have
\bea
g_1 & = & -2  \frac{m_{W}^2}{m_H^2}Y_{ee} Y_{e\mu}^* \,, \ \   g_2  =  -2 \frac{m_{W}^2}{m_H^2} Y_{ee}Y_{\mu e}\,, \ \
g_3  =  \sum_{V^0} \frac{m_{W}^2}{m_{V^0}^2} D_R^{ee}D_R^{e\mu}\,, \ \ 
g_4  =  \sum_{V^0}  \frac{m_{W}^2}{m_{V^0}^2} D_L^{ee}D_L^{e\mu} \,, \nn  \\
g_5 & = & \sum_{V^0} \frac{m_{W}^2}{m_{V^0}^2} D_L^{ee}D_R^{e\mu}+\frac{m_{W}^2}{m_H^2}Y_{ee}Y_{e\mu }^* \,,  \hspace{1.2cm}
g_6  =  \sum_{V^0} \frac{m_{W}^2}{m_{V^0}^2} D_R^{ee}D_L^{e\mu}+\frac{m_{W}^2}{m_H^2}Y_{ee}Y_{\mu e}, \label{g1g6}
\eea
where $m_W$ and $m_H$ are the masses of the SM $W$ and Higgs bosons, respectively. 
The couplings in  (\ref{Lagquark}) are given by
\be
g_{LV(q)}= 4\sum_{V^0} \frac{m_{W}^2}{m_{V^0}^2}g_q D_L^{e\mu}\,, \ \ \ \ 
g_{RV(q)}= 4\sum_{V^0}\frac{m_{W}^2}{m_{V^0}^2}g_q D_R^{e\mu} \,.
\label{gLVqgRVq}
\ee
Strictly speaking, the effective couplings appearing in (\ref{LeffFCNC}) and (\ref{Lagquark}) should be evaluated at the scale of the decaying charged lepton mass, while
(\ref{ALAReff}), (\ref{g1g6}) and (\ref{gLVqgRVq}) give the couplings at the energy scale corresponding to the mass of the state
that has been integrated out. Contrary to, say, non-leptonic quark decays,
renormalization group effects in leptonic decays are sub-leading and can be neglected in first approximation. We can then directly identify  the coefficients (\ref{ALAReff}), (\ref{g1g6}) and (\ref{gLVqgRVq})
as the low-energy couplings relevant for the LFV processes.

We have numerically computed the LFV processes by keeping, for each independent KK tower of states, the first
heavy KK mode.  For tree-level processes this approximation is quite accurate and should differ from the full result by ${\cal O}(10\%)$, as we have numerically checked by keeping more KK states. For radiative decays, the approximation is less accurate and might differ from the full result by ${\cal O}(50\%)$. This accuracy is enough for our purposes. If one demands a higher precision, a full 5D computation, as e.g. in \cite{Csaki:2010aj}, should be performed, although one should keep in mind that the limited range of validity of the effective field theory of 5D warped models puts a stringent bound on the accuracy one can in principle achieve.

As we already said, all LFV processes are induced by the BKT (\ref{LBKT}). More precisely, LFV processes
are induced by the relative differences in the $z_{el}$, $z_{\mu l}$, $z_{\tau l}$ factors, since universal BKT 
simply amount to a trivial rescaling of the fields. Let us first give an estimate of the relative relevance of the couplings $g_1$--$g_6$ and $g_{L/R V(q)}$. They are induced by the tree-level exchange of Higgs and neutral gauge bosons, namely the SM boson $Z$ and its first KK mode $Z^{(1)}$, 
the first KK mode of the photon $\gamma^{(1)}$, the first KK mode of the neutral $SO(5)/SO(4)$ fields $A^{\hat 3 (1)}$
and $A^{\hat 4 (1)}$,  the first KK mode of the 5D gauge field $Z^{\prime (1)}$.
The 5D fields $Z$, $\gamma$ and $Z^\prime$ are related as follows to the $SO(5)\times U(1)_X$ fields
$W_{3L}$, $W_{3R}$ and $X$:
\bea
B  &= & \frac{g_{5X}W_{3R}+g_5 X}{\sqrt{g_5^2+g_{5X}^2}}\,, \hspace{1.2cm} Z^\prime = \frac{g_{5}W_{3R}-g_{5X} X}{\sqrt{g_5^2+g_{5X}^2}}\,, \nn \\
Z &  = & \cos \theta_W W_{3L} - \sin\theta_W B\,, \ \ \gamma = \cos\theta_W B + \sin\theta_W W_{3L} \,, 
\eea
with $g_{5X}$ the 5D coupling of the $U(1)_X$ field, determined in terms of $\theta_W$:
\be
\tan^2\theta_W = \frac{g_{5X}^2}{g_5^2+g_{5X}^2}\,.
\ee
Due to the IR-peaked profile of the KK wave functions, the leading effect of (\ref{LBKT}) is to mix the LH zero mode fields $l_{\alpha L}^{(0)}$ among themselves.  The main source of flavour violation clearly arises from LH fields. Since fermion Yukawa couplings are negligible, we have
\be
g_1\simeq g_2 \simeq g_3 \simeq g_5 \simeq g_{RV(q)} \simeq 0\,.
\ee
The LFV couplings $D_L^{e\mu}$ in (\ref{IntLagLFV}) govern the size of the relevant effective couplings  $g_4$, $g_6$ and $g_{LV(q)}$.
 The dominant LFV effects arise from the rotation and rescaling of $l_{\alpha L}^{(0)}$ necessary to get canonically normalized kinetic terms.
Before EWSB effects are considered, no flavour violation is expected from the SM $Z$ boson by gauge invariance. 
The leading deviations arise from the gauge fields 
$Z^{(1)}$, $\gamma^{(1)}$ and $Z^{\prime (1)}$. It is straightforward to derive a reasonable accurate estimate for the couplings $D_L^{e\mu}$: 
\bea
D_L^{e \mu}(Z^{(1)}) & \simeq &  (g^{Z^{(1)}}_{loc}-g^{Z^{(1)}}_{bulk}) ( Z_l)_{e \mu} \,, \nn \\
D_L^{e \mu}(\gamma^{(1)}) & \simeq &  (g^{\gamma^{(1)}}_{loc}-g^{\gamma^{(1)}}_{bulk}) ( Z_l)_{e \mu}
\,, \nn \\
D_L^{e \mu}(Z^{\prime (1)}) & \simeq & -g^{Z^{\prime(1)}}_{bulk} (Z_l)_{e \mu} 
\,, 
\label{DLestimate}
\eea
where $g_{loc}$ and $g_{bulk}$ are the BKT and bulk contributions to the gauge couplings, respectively.
When EWSB effects are considered, LFV effects are transmitted to the SM $Z$ boson as well. The resulting $D_L^{e\mu}(Z)$ is suppressed by the mixing, but the latter is approximately compensated by the absence of the mass suppression factors appearing in the couplings $g_i$ (\ref{g1g6}).
Eventually, the SM $Z$ boson contribution to LFV is of the same order of magnitude of that of the fields $Z^{(1)}$, $\gamma^{(1)}$ and $Z^{\prime(1)}$. In (\ref{DLestimate}), $Z_l$ is the effective BKT felt by the zero mode, which is obtained by multiplying $\hat Z_l$ by the square of the zero-mode wave function  (\ref{LexpKK0}) evaluated at the UV brane:
\be
(Z_l)_{\alpha\beta} = \Big(\frac{R}{R^\prime}\Big)^{1-2c_l}f_{c_l}^2 \frac{1}{\sqrt{\rho_\alpha}} (\hat Z_l)_{\alpha\beta}  \frac{1}{\sqrt{\rho_\beta}}\,.
\label{BKTeff}
\ee
For $c_l>1/2$, the factor entering in (\ref{BKTeff}) becomes of ${\cal O}(1)$, while it is exponentially small for $c_l<1/2$. 
For the relevant region where $c_l\simeq 1/2$ and $\rho_\alpha\simeq 1$, the effective BKT $Z_l$ is considerably smaller than $\hat Z_l$. For $c_l=1/2+\delta$, at linear order in $\delta$, we have
\be
Z_l \simeq \Big(\log^{-1}\frac{R^\prime}{R}+\delta \Big)\hat Z_l \simeq  \Big(\frac{1}{35}+ \delta\Big) \hat Z_l \,.
\label{BKTeff2}
\ee
The effect of the BKT on the LFV is naturally suppressed. {\it This is the main reason why most of the parameter space of our model
successfully passes the bounds imposed by LFV processes}. The suppression factor (\ref{BKTeff2}) also explains why the approximation of treating the  BKT as insertions is valid
even for ${\cal O}(1)$ BKT at the UV brane. 

\begin{figure}[t!]
\begin{minipage}[t]{0.465\linewidth} 
\begin{center}
\includegraphics*[width=\textwidth]{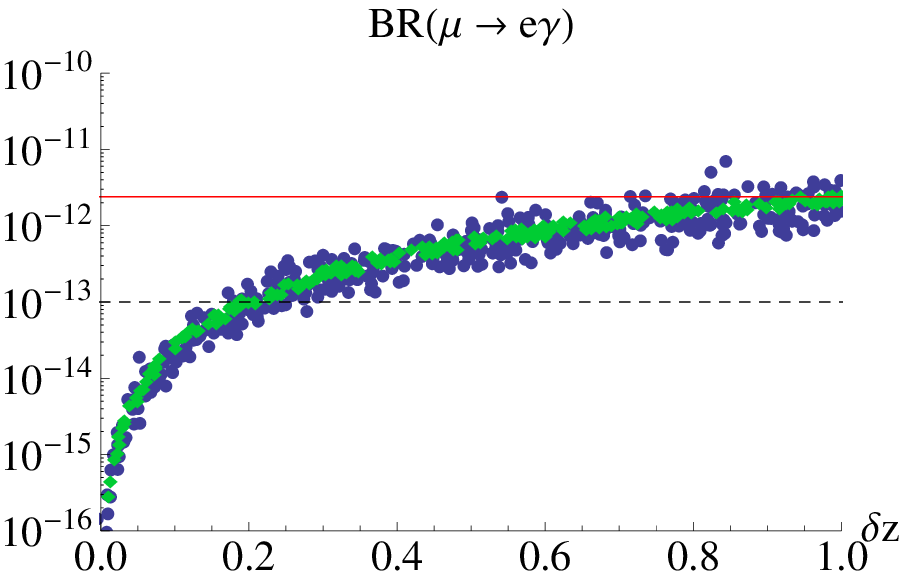}
\end{center}
\end{minipage}
\hspace{0.5cm} 
\begin{minipage}[t]{0.48\linewidth}
\begin{center}
\includegraphics*[width=\textwidth]{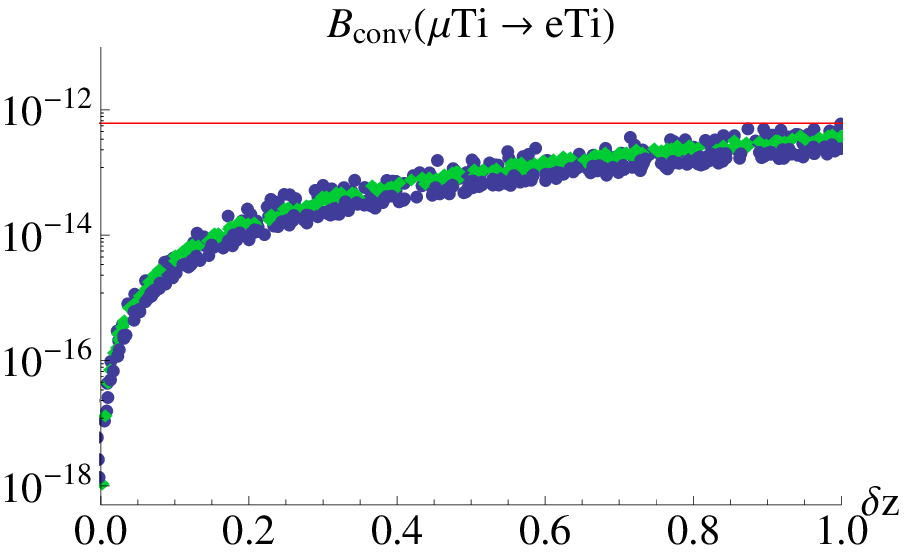}
\end{center}
\end{minipage}
\caption{\small Branching ratio of $\mu\rightarrow e \gamma$ and $\mu-e$ conversion in Ti as a function of the UV BKT $\delta z\equiv z_{\mu l}-z_{e l}$. 
The continuous (red) and dashed (black) lines in the left panel are the actual \cite{MEGactual} and the expected future bound \cite{MEG} given by the MEG 
experiment. The (red) line in the right panel is the current bound as given by the SINDRUM II experiment \cite{Sindrum2}. The plots refer to the Majorana model with $m_{{\rm IR},\alpha}^\nu=1$, $c_l=0.52$, $c_\nu = -0.365$, $h=1/3$ and normal neutrino mass hierarchy. The IR masses $m_{{\rm IR},\alpha}^l$ are random numbers chosen
between $0.05$ and $1.5$ for $m_{{\rm IR},e,\mu}^l$ and $0.5$ and $1.5$ for $m_{{\rm IR},\tau}^l$
(blue points) or all set to one (green diamonds).
The masses $m_{\mathrm{UV},\alpha}$ are chosen such that  
 the lightest neutrino mass is $m_0=0.01$ eV and the best fit values \cite{nudata} of the solar and atmospheric mass square differences $\Delta m_{\rm{sol}}^2 = 7.59 \times 10^{-5}$ eV$^2$ and 
$\Delta m_{\rm{atm}}^2 = 2.40 \times 10^{-3}$ eV$^2$ are reproduced using (\ref{neutrinomassZMA}), corrected for the effect of the BKT.}
\label{fig:mue-mueconv-M}
\end{figure}

Let us now consider the couplings $A_L$ and $A_R$. It is immediately clear from the more composite nature of the
muon with respect to the electron that $A_L\ll A_R$, so in first approximation $A_L$ can be neglected. 
Higgs and neutral gauge boson mediated contributions $A_R^{(H)}$ and $A_R^{(Z)}$ are also negligible, and the dominant contribution $A_R^{(W)}$ arises from the charged gauge bosons with Neumann b.c. at the UV brane, namely the SM $W$ boson and its first KK excitation $W_L^{(1)}$. It turns out to be rather difficult to derive an accurate analytic formula for $A_R^{(W)}$ since neutrino, charged lepton and gauge boson Yukawa couplings significantly contribute to the branching ratio. An order of magnitude estimate can be obtained by focusing on a definite contribution that is always one of the dominant ones, although not the only one. It arises from the Yukawa couplings between the SM neutrinos $l_L^{u(0)}$ and the 
RH singlet fields $N^l_{\alpha R}$,  the combination of $N^{(0)}_{\alpha R}$ and $N^{(1)}_{\alpha R}$ orthonormal to the heavy Majorana fermions $N^h_{\alpha R}$. It is relevant because these Yukawas are sizable and
$N^l_\alpha$ is typically the lightest fermion resonance in the model.  We get
\be
A_R^{(W)} \sim \frac{i c}{16\pi^2} \Big(\frac{Y}{m_{N_l}}\Big)^2 (Z_l)_{e \mu}\,,
\label{ARWana}
\ee
where $c$ is an order $1$ coefficient and $Y$ is the approximate flavour universal value of the Yuakwa coupling in the original basis of fields, before the redefinitions
needed to get canonically normalized kinetic terms.  We plot in figure \ref{fig:mue-mueconv-M} the bounds arising from $\mu\rightarrow e \gamma$ 
and $\mu-e$ conversion in Ti (the most constraining case) as a function of $\delta z \equiv z_{\mu l}-z_{e l}=3(\hat Z_l)_{e\mu}$.
Both processes depend quadratically on $\delta z$, as expected from (\ref{DLestimate}) and (\ref{ARWana}). As can be seen from figure \ref{fig:mue-mueconv-M},  the IR masses $m_{\rm{IR},\alpha}^l$ do not play an important role, provided that $m_{\rm{IR},\tau}^l$ is large enough, as required by $\delta g^\tau_{l_R}$.
Thanks again to the suppression factor appearing in (\ref{BKTeff2}),  the branching ratio is almost always below the current limit of $2.4 \times 10^{-12}$ for $|\delta z|<1$. Using however the future bound expected from the MEG experiment of $10^{-13}$, we find that $|\delta z|$ is constrained to be less than 0.25.
The decay to three leptons $\mu\rightarrow 3e$ and radiative $\tau$ decays are always well below the experimental bounds and
are not reported.\footnote{Notice that, due to the smallness of the couplings of the SM leptons to the neutral KK gauge bosons for $c_l \gtrsim 1/2$, the contribution of $A_R^{(W)}$ to (\ref{BReff}) is comparable to that given by the couplings $g_4$ and $g_6$.} 
We also performed an analysis for larger $m_0$ and for both, normal and inverted, neutrino mass orderings,  with results identical to those
shown in figure 1.

Let us finally consider the bounds arising from lepton mixing, assuming vanishing phases. 
As we have already mentioned, in order to avoid too large deviations 
from TB lepton mixing, the IR localized neutrino mass terms $m_{{\rm IR},\alpha}^\nu$ should be taken close to universal.
Parametrizing $m_{\rm{IR},\alpha}^\nu$ in the following way: $m_{{\rm IR},e}^\nu=m_{{\rm IR}}^\nu$,  $m_{{\rm IR},\mu}^\nu=m_{{\rm IR}}^\nu(1+\delta m_{{\rm IR}}^\nu)$, 
$m_{{\rm IR},\tau}^\nu=m_{{\rm IR}}^\nu$,\footnote{We have chosen this particular parametrization of $m_{\rm{IR},\alpha}^\nu$, since in this way all mixing
angles are subject to a deviation linear in $\delta m_{\rm{IR}}^\nu$ and neither accidental cancellation nor accidental enhancement of the coefficient of the linear
perturbation is encountered.} we can analyze neutrino masses and mixing arising from the light neutrino mass matrix in (\ref{neutrinomassZMA}) in an expansion in $\delta m_{\rm{IR}}^\nu$. 
We neglect the effects of BKT in the following and take $c_l=0.52$ and $h=1/3$ so that the parameters $\rho_\alpha$ are universal to a good approximation.
For normally ordered light neutrinos with $m_0=0.01$ eV and mass
square differences given by the experimental best fit values \cite{nudata}, the mixing angles turn out to be
\bea\nonumber
&&\sin \theta_{13} \approx 0.05 \, |\delta m_{\rm IR}^\nu|\, ,\\
\nonumber 
&& \sin^2 \theta_{23} \approx \frac{1}{2} + 0.82 \, \delta m_{\rm IR}^\nu\, ,\\
&& \sin^2 \theta_{12} \approx \frac{1}{3} - 1.58 \, \delta m_{\rm IR}^\nu \, ,
\eea
showing that the requirement of having $\sin^2 \theta_{12}$ in the experimentally allowed $3 \sigma$ range \cite{nudata}, $0.27 \lesssim \sin^2 \theta_{12} \lesssim 0.38$,
leads to the constraint
\be
-0.03 \lesssim \delta m_{{\rm IR}}^\nu \lesssim 0.04\, .
\ee
At the same time $\sin^2 \theta_{23}$ remains within its $1 \sigma$ range, $0.475 \lesssim \sin^2 \theta_{23} \lesssim 0.533$. The reactor
mixing angle $\sin^2 \theta_{13}$ takes as maximal value $4 \times 10^{-6}$, well below the current and prospective future bounds.
These statements are in agreement with our numerical results.
The validity of the expansion in $\delta m_{\rm IR}^\nu$ strongly depends on $m_0$. For instance, by taking 
 $\delta m_{{\rm IR}}^\nu=0.1$, the perturbative expansion in $\delta m_{\rm IR}^\nu$ breaks down for $m_0\gtrsim 0.03$ eV.

We also performed a study for inverted mass hierarchy. In this case the above perturbative expansion is not valid for any value of $m_0$.
From the numerical results we see that large corrections to the solar mixing angle always arise, whereas the atmospheric mixing angle 
gets small corrections and the reactor mixing angle remains always very small.
The large deviations of $\theta_{12}$ can be easily understood by noticing that for inverted neutrino mass ordering the relative splitting between 
the two heavier light neutrinos is in general small compared to the scale $\sqrt{m_0^2 + \Delta m_{\rm{atm}}^2} \gtrsim 0.049$ eV. Thus, the angle $\theta_{12}$ associated with the mixing in this almost degenerate
sub-sector is subject to large deviations from its initial TB value even for very small deviations $\delta m_{\rm{IR}}^\nu$ from universality.
As a consequence, the latter have to be as small as possible in the case of inverted neutrino mass ordering, which is most naturally achieved in the $\Z_2$-invariant model.

In summary, in the case of normally ordered light neutrinos and a rather small mass scale $m_0$, deviations from universality of $m_{\rm{IR},\alpha}^\nu$ are
admissible up to the level $|\delta m_{\rm{IR}}^\nu| \lesssim 0.04$. Generically, the solar mixing angle, which is the most precisely measured one up to date in neutrino oscillation
experiments, turns out to be the most sensitive one to corrections. For a neutrino mass spectrum with inverted hierarchy, the most natural situation is the one in which
an additional accidental $\Z_2$ exchange symmetry on the IR brane renders the mass terms $m_{\rm{IR},\alpha}^\nu$ universal.
The deviations of $\theta_{12}$ and $\theta_{23}$  are well under control in the $\Z_2$-invariant model for all values of $m_0$ and
both types of neutrino mass hierarchy (with $\sin^2 \theta_{23}$ in the experimentally allowed $1\sigma$ range and $\sin^2 \theta_{12}$ in the $2\sigma$ range). 
The angle $\theta_{13}$  is in this case always constrained to be very small, $\sin ^2 \theta_{13} \lesssim 10^{-6}$, and 
cannot be detected.

\subsection{Uncalculable Corrections and $\tau$ Decays}
\label{subsec:tauUV}

Contrary to the operators in (\ref{LeffFCNC}), where only two flavours appear, LFV operators
involving simultaneously three different flavours are not constrained effectively by our choice of discrete symmetries. Dimension 8, 4 fermion $S_4\times \Z_3$ ($\Z_3^{(D)}$) invariant bulk (IR localized) operators reducing to flavour violating dimension 6 LL, RR and LR/RL operators can be constructed.
Among these, the ones of the form $(\bar \tau\Gamma \mu) (\bar e\Gamma  \mu)$,  $(\bar e\Gamma  \mu) (\bar e \Gamma \tau)$, with $\Gamma = \gamma^\mu, \gamma^5$, and their hermitian conjugates, can directly mediate the $\tau$ decays $\tau\rightarrow e2\mu$ and $\tau\rightarrow \mu 2e$. 
The branching ratio for these decays is of order $10^{-8}$ \cite{pdg}. 
The size of the couplings of these operators, uncalculable within the 5D theory, can be estimated by using na\"ive dimensional analysis.
For IR brane operators (bulk operators give roughly the same result) we get
\be
\kappa_{UV} g_5^2 R^3 \Big(\frac{R}{R^\prime}\Big)^4 \bar \xi \xi \bar\xi \xi (z=R^\prime) \simeq \kappa_{UV} g^2  (R^\prime)^2\log\Big(\frac{R^\prime}{R}\Big) \bar l^{(0)} l^{(0)} \bar l^{(0)} l^{(0)} \prod_{n=1}^4 f_{c_n}\,,
\ee
where $c_n = -c_\alpha$ for RH leptons, $c_n = c_l$ for LH leptons, $\kappa_{UV}$ is an ${\cal O}(1)$  dimensionless coupling
and in the second equality we have plugged  in the zero mode wave function of the SM leptons $l^{(0)}$ for the 5D fermion fields $\xi$. 
The most stringent bounds arise from the LL operators, since $f_{-c_e}, f_{-c_\mu}\ll f_{c_l}$.  By matching with (\ref{LeffFCNC}), we get
\be 
g_{UV}^{eff} \simeq 2 \kappa_{UV} (m_W R^\prime)^2\log\Big(\frac{R^\prime}{R}\Big)  \prod_{n=1}^4 f_{c_n}\,.
\ee
We demand that $g_{UV}^{eff} \lesssim 2\times 10^{-4}$.\footnote{Notice that in $\tau$ decays, the formula (\ref{BReff}) gets a suppression factor $\simeq 0.18$ that takes into account of the hadronic contribution to the total decay width.  This explains the factor 2 in $2\times 10^{-4}$.} 
For $\kappa_{UV} \sim{\cal O}(1)$, $1/R^\prime \gtrsim 1.5$ TeV and $1/R\simeq M_{Pl}$, we get the following lower bound for $c_l$:
\be 
c_l \gtrsim 0.5\,.
\label{clUVbound}
\ee
Notice that flavour preserving dimension 8 operators are also potentially dangerous, contributing, e.g., to the deviation from the SM values of the couplings of leptons to the vector bosons.
From a quick estimate, we find that the bound  (\ref{clUVbound}) is more constraining. Summarizing, demanding that the uncalculable contributions coming from higher dimensional operators are sufficiently suppressed results in a bound on the degree of compositeness of the SM leptons.

%%%%%%%%%%%%%%%%%%%%%%%%%%%%%%%%%%%%%%%%%%%%
\section{Explicit 5D Dirac Model}
\label{sec:leptonDirac}
%%%%%%%%%%%%%%%%%%%%%%%%%%%%%%%%%%%%%%%%%%%%

In this section we provide an explicit 5D gauge-Higgs unification warped model realizing the Dirac scenario outlined in subsection \ref{subsec:Dscenario}.
The model is very closely related to the Majorana model of section \ref{sec:leptonM}, so we focus on the key differences between the two. 
The gauge symmetry and its breaking pattern is the same as before, while the flavour symmetry is slightly different:
\be
G_{\mathrm{flavour}}=S_4\times \Z_3\times \Z_5 \times  \Z^\prime_3 \,,
\ee
broken to $G_{\mathrm{flavour, UV}}=\Z_2\times \Z_2\times\Z_3  \times  \Z_3^\prime$ and 
$G_{\mathrm{flavour, IR}}=\Z^{(D)}_3\times \Z_5$ at the UV and IR branes, 
respectively. 
Like in the Majorana model, $\Z_5$ and $\Z_3^\prime$ are included  to constrain the number of terms allowed at the UV and IR boundaries.

The particle content and b.c. for the fields are identical to those in the Majorana model, with the only exception of a crucial flip in the b.c. for the singlet  neutrino $\hat \nu$
in the ${\bf 5}$:
\be
\hat{\nu}_{\alpha L} \, (-+) \rightarrow \hat{\nu}_{\alpha L} \, (+-)\,.
\ee
The flavour properties of the fields  are summarized in table \ref{table:flavourD}. Notice that the discrete symmetries forbid the appearance
of any bulk or boundary Majorana mass term.

\begin{table}
\begin{tabular}{|c|c|c|c|}\hline
& Bulk & UV & IR\\[0.05in]
\cline{2-4}
 & $G_{\mathrm{flavour}}=S_4\times \Z_3\times \Z_5\times \Z_3^\prime$ & $G_{\mathrm{flavour,UV}}=\Z_2\times \Z_2\times \Z_3 \times \Z_3^\prime$ & \rule[0.1in]{0cm}{0cm} $G_{\mathrm{flavour,IR}}=\Z_3^{(D)}\times \Z_5$\\[0.05in]
\hline
 &   & \rule[0.05in]{0cm}{0cm} $(1,-1,1,\omega)$ & \\
 $\xi_{l,\alpha}$         &  $({\bf 3}, 1, \omega_5,\omega)$                                 & $(-1,1,1,\omega)$ & $(\omega^{2 (\alpha-1)},\omega_5)$\\
          &                                   & $(-1,-1,1,\omega)$ & \\[0.05in]
\hline
\rule[0.1in]{0cm}{0cm}$\xi_{e,\alpha}$ &  $({\bf 1}, \omega^{2 (\alpha-1)}, \omega_5, \omega)$ & $(1,1,\omega^{2 (\alpha-1)},\omega)$ & $(\omega^{2 (\alpha-1)},\omega_5)$\\[0.05in]
\hline
 &   & \rule[0.05in]{0cm}{0cm} $(1,-1,1,\omega)$ & \\
 $\xi_{\nu,\alpha}$         &  $({\bf 3}, 1, \omega_5^2,\omega)$                                 & $(-1,1,1,\omega)$ & $(\omega^{2 (\alpha-1)},\omega_5^2)$\\
          &                                   & $(-1,-1,1,\omega)$ & \\[0.05in]
\hline
\end{tabular}
\caption{Transformation properties of the 5D multiplets $\xi_l$, $\xi_e$ and $\xi_\nu$ under $G_{{\rm flavour}}$ and their decomposition properties under the subgroups $G_{\mathrm{flavour, UV}}$ and $G_{\mathrm{flavour, IR}}$ in the Dirac model. $\omega_5$ is the fifth root of unity $\omega_5 \equiv \mathrm{e}^{2 \pi i/5}$.}
\label{table:flavourD}
\end{table}

The invariant mass terms at the IR and UV branes are
\bea
-\mathcal{L}_{\mathrm{IR}} & = & \left(\frac{R}{R'}\right)^4 \sum \limits_{\alpha=e,\mu,\tau}^3 m_{\mathrm{IR},\alpha}^l 
\left(\ol{\wt{L}}_{1,\alpha L} \wt{L}_{2,\alpha R} + \ol{L}_{\alpha L} \hat{L}_{\alpha R}\right)
+ h.c.  \nn \\
-\mathcal{L}_{\mathrm{UV}} &=& \ol{\hat{\nu}}_{\alpha L} \mathcal{M}_{\mathrm{UV},\alpha\beta} \nu_{\beta R} + h.c.
\eea
with $\mathcal{M}_{\mathrm{UV}}$ as in (\ref{MUV}). 
The phases of the IR masses $m_{{\rm IR},\alpha}^l$ can still be absorbed by re-defining the 5D $SO(5)$ fields
$\xi_{e,\alpha}$ and one of the three phases contained in $\mathcal{M}_{{\rm UV}}$ through re-phasing the fields $\xi_{\nu,\alpha}$. Again, we
are left with two non-trivial phases coming from the UV mass terms.

\subsection{Mass Spectrum}

The KK expansion in the ZMA of the doublets $L_{\alpha L}$, $\hat L_{\alpha L}$ and $e_{\alpha R}$ is identical to (\ref{LexpKK0}), (\ref{LhatexpKK0}) and (\ref{eRexpKK0}) and gives rise to the same charged lepton mass matrix (\ref{CLMassMatrix}). 

The KK expansion of neutrinos is of course  different. The IR b.c. are not affected by mass terms, while the UV b.c. read
\be
\hat{\nu}_{\alpha R} = \mathcal{M}_{\mathrm{UV},\alpha\beta}\nu_{\beta R} \,, \ \ \ \ \ \ \ 
\nu_{\alpha L} = -\mathcal{M}_{\mathrm{UV},\beta\alpha}^* \hat{\nu}_{\beta L}
\ee
and lead to the following canonically normalized zero mode expansion
\begin{eqnarray}
&& \nu_{\alpha R}(x,z) = \frac{1}{\sqrt{R'}} \left( \frac{z}{R} \right)^2  \left( \frac{z}{R'} \right)^{c_\nu} f_{-c_\nu} (U_{TB})_{\alpha\beta} \frac{1}{\sqrt{\kappa_\beta}} N_{\beta R}^{(0)} (x) \,, \nn
\\
&& \hat{\nu}_{\alpha R}(x,z) = \frac{1}{\sqrt{R'}} \left( \frac{z}{R} \right)^2  \left( \frac{z}{R'} \right)^{c_l} \left( \frac{R}{R'} \right)^{c_\nu-c_l}
f_{-c_\nu} (U_{TB})_{\alpha\beta} \frac{m_{\mathrm{UV,\beta}}}{\sqrt{\kappa_\beta}} N_{\beta R}^{(0)} (x)\,,
\label{nuDKK}
\end{eqnarray}
where 
\be
\kappa_\alpha = 1 + |m_{\mathrm{UV},\alpha}|^2 \left( \frac{f_{-c_\nu}}{f_{-c_l}} \right)^2 
\left( \frac{R}{R'} \right)^{2 (c_\nu -c_l)}.
\label{kappa}
\ee
By computing the wave function overlap with the Higgs field, we get the neutrino mass matrix:
\begin{equation}
M_{\nu, \alpha\beta} = \frac{h}{\sqrt{2}} f_{c_l} f_{-c_\nu} \left( \frac{R}{R'} \right)^{c_\nu+\frac 12 -(c_l-\frac 12)} \frac{1}{\sqrt{\rho_\alpha}} (U_{TB})_{\alpha\beta} \frac{m_{\mathrm{UV},\beta}}{R} \frac{1}{\sqrt{\kappa_\beta}}\,,
\label{massneutrinoD}
\end{equation}
where $\rho_\alpha$ is defined as in (\ref{rhogammaDEF}). Thanks to the factor $(R/R^\prime)$ in (\ref{massneutrinoD}),  the correct order of magnitude for neutrino masses is naturally obtained by choosing 
\be
c_\nu - c_l  \approx 0.8\,.
\ee
The only source of deviation from TB mixing in (\ref{massneutrinoD}) is given by the factor $\rho_\alpha$, which should be contrasted with the situation in the Majorana model,
where the deviations are given by $\rho_\alpha$ and the neutrino mass terms $m_{\mathrm{IR},\alpha}^\nu$. 

Let us consider the relation between the 5D model and the general 4D analysis performed in subsection \ref{subsec:Dscenario}. The anomalous dimensions of $\Psi^\alpha_{l_L,R}$, $\Psi^\alpha_{l_R,L}$ are the same as in the Majorana model.
Since $\hat\nu_L$ and $\nu_L$ belong to the same 5D bulk multiplet, we have $\gamma_{\hat\nu L} =\gamma_{lL}$. The states denoted by $\hat\nu^\alpha$ in (\ref{masshatnu})  are the lightest KK vector-like states of the tower of modes coming from $\hat \nu_{\alpha }$ and $\nu_\alpha$ in the 5D model. Their masses are determined as the zeros of a certain combination of Bessel functions and are approximately flavour independent. For $c_l\gtrsim 0.44$, we have 
 \be
 (m_{\hat \nu}^\alpha R^\prime)^2 \simeq \frac{1-4c_l^2}{(c_l+\frac 12)- \big(\frac{R^\prime}{R}\big)^{2c_l-1}\Big(1+\frac{|m_{\rm{UV},\alpha}|^2(1-2c_l)}{\Gamma(c_\nu-1/2)(1-2c_\nu)}\Big)}\,.
 \label{mNana}
  \ee
The coefficients $d_\alpha$ appearing in (\ref{masshatnu}) are correspondingly flavour independent in first approximation. Along the lines of \cite{Contino:2004vy}, the parameter $\epsilon$ defined in subsection \ref{subsec:Dscenario} can be seen to arise from the mixing between two heavy elementary fermions $\Psi_L$ and $\Psi_R$ of opposite chiralities  with a RH massless bound state $\nu_R$ of the CFT, all in ${\bf 3}$ of $S_4$. Omitting flavour indices, the relevant Lagrangian is 
\be 
{\cal L} = \bar \Psi (i \hat\partial - \Lambda) \Psi + \frac{c} {\Lambda^{\gamma_{\nu_R}}}(\bar \Psi_L \nu_R+\bar \nu_R \Psi_L) + \big(\bar{\hat\nu}_L M \Psi_R + h.c.\big) 
\ee 
with $\gamma_{\nu R} = |c_\nu+1/2|-1$ the anomalous dimension of $\nu_R$ and $c$ a universal ${\cal O}(1)$ coefficient. If we assume that $\Lambda$ is flavour independent, all the non-trivial flavour dependence is in the mass term $M$, which is of the form (\ref{MUTB}). Integrating out the fields $\Psi$ gives at leading order 
\be 
\Psi_R\simeq \frac{c}{\Lambda^{\gamma_{\nu R}+1}} \nu_R\,. 
\ee 
Rescaling $\nu_R\rightarrow \mu^{\gamma_{\nu R}+1}\nu_R$ to effectively get canonical kinetic terms for a free fermion field, gives 
\be 
\epsilon \sim  c \Big(\frac{R}{R^\prime}\Big)^{|c_\nu+1/2|} \,. 
\label{epsilon} 
\ee 
The form of the coefficients $\hat d_\alpha$ introduced in (\ref{nuhatINT}) will be determined in subsection \ref{subsec:devD}. We anticipate here that they
are flavour independent, coming from $S_4$ invariant bulk interactions.  It is important to notice that for $\gamma_{\hat\nu  L} =\gamma_{l L}=c_l-1/2$ (taking $c_l+1/2>0$),  (\ref{nuhatINT})  shows a non-decoupling effect. For $\gamma_{\hat\nu L}>0$ the explicit $(\mu/\Lambda)$ suppression factor in (\ref{nuhatINT}) cancels the one coming
from $m_{\hat \nu}$. For $\gamma_{\hat\nu  L}<0$, the mass of $\hat \nu$ is unsuppressed, but the LH leptons are mostly composite and one has to perform a field rescaling to get the canonically normalized LH field $\nu_L$, 
as  in (\ref{nuKinComp}).  Its effect is again to compensate for the explicit factor $(\mu/\Lambda)$ in (\ref{nuhatINT}).  For any  $\gamma_{\hat\nu  L}$, then, we get $\hat\nu_L\sim {\cal O}(h) \nu_L$, a result that leads
to unsuppressed deviations of neutrino couplings to the $W$ and $Z$ from their SM values, as shown below.  The non-trivial factor  $\kappa_\alpha$ in  (\ref{massneutrinoD})  comes in the 4D picture from corrections to the kinetic term of $\nu_R$ we have neglected, appearing when the heavy fields $\Psi$ are integrated out.
They are completely negligible, given the suppression factor appearing in (\ref{kappa}).  The factors $\rho_\alpha$, as in the Majorana model, encode corrections to the kinetic term of $l_L^{(0)}$ coming from the composite sector.
Summarizing, the mass formula (\ref{massneutrinoD}) is a particular realization of the more general expression (\ref{massDirac4d}) where all deviations from TB mixing are naturally suppressed.

In the limit of an $S_4$ invariant IR Lagrangian, $m_{\mathrm{IR},\alpha}^l\rightarrow 0$, 
the neutrino mass matrix (\ref{massneutrinoD}) leads to exact TB mixing. However, the factors $\rho_\alpha$ disfavour composite LH leptons, because the more these states are composite, the smaller $m_{\mathrm{IR},\alpha}^l$ should be 
to keep $\rho_\alpha \simeq 1$. Bounds on gauge coupling deviations favour the region in parameter space where $c_l \lesssim 1/2$. Given that TB lepton mixing  and (\ref{clUVbound}) favour $c_l\gtrsim 1/2$, the region $c_l\simeq 1/2$ is again the one of interest.

The mass spectrum of the neutral KK resonances in the Dirac model differs from that in the Majorana model mostly for the presence of the light states $\hat\nu$. 
For the benchmark values $h=1/3$, $c_l=0.52$, (\ref{mNana}) gives $m_{\hat \nu}^\alpha\gtrsim 200$ GeV.
The masses of the next-to-lightest charged and neutral KK fermion resonances (taking $m_{{\rm IR},\alpha}^l$ between $1/2$ and $3/2$)
are slightly below 2 TeV, so approximately comparable to the spectrum found in the Majorana model.
The KK gauge boson masses are obviously identical in the two cases.

\subsection{Deviations from Gauge Coupling Universality}

\label{subsec:devD}

The realization of the SM charged leptons in the 5D Dirac and Majorana models is identical, so (\ref{deltagL}) and (\ref{deltagR2}) continue to apply.
In the Dirac model, $\nu_{\alpha L}$ are still embedded into $SO(5)$ multiplets with $T_{3L}\neq T_{3R}$, so non-trivial deviations are expected.
The holographic analysis is complicated by the presence of the 4D singlet fields $\hat \nu_{\alpha L}(x,z=R)$ that should be kept and eventually integrated out.\footnote{Recall that
in the holographic approach the 5D fields evaluated at the UV brane can directly be identified with the fields in the elementary sector defined in section 2.}  
Omitting intermediate steps, one simply gets
\be
\hat \nu_{\alpha L} (x,z=R) = -\Big(\frac{1}{\sqrt{2}} \tan h \Big)\nu_{\alpha L} (x,z=R) \,,
\label{nuhatnu}
\ee  
independently of any parameter. As anticipated below (\ref{epsilon}), a non-decoupling occurs in 
(\ref{nuhatINT}).\footnote{A similar non-decoupling effect has recently been noted in 5D models in flat space, see (3.15) of \cite{Panico:2010is}.} At leading order in $h$, (\ref{nuhatnu}) leads to the following universal deviation
\be
\frac{\delta g^\alpha_{\nu_L}}{g^\alpha_{\nu_L}} =  - \frac{h^2}{2} \,, \ \ \ 
\left|\frac{\delta g^\alpha_{\nu_L l_L}}{g^\alpha_{\nu_L l_L}}\right| = \frac 12 \left|\frac{\delta g^\alpha_{\nu_L}}{g^\alpha_{\nu_L}}\right|  \,.
\label{gnuLD}
\ee
By demanding (\ref{EWPbounds}), we get the universal bound 
\be
h \lesssim \frac{1}{10}
\label{alphabound}
\ee
independently of $c_l$. In light of the bound (\ref{alphabound}), the Dirac model appears to be fine-tuned at ${\cal O}(1\%)$ level, unless one advocates exotic hidden physics
that is responsible for a fraction of the invisible partial width of the $Z$ boson. 

\subsection{LFV Processes and BKT}

\label{subsec:FboundsD}

Several considerations made in the Majorana model continue to apply in the Dirac case.
The form of the interaction Lagrangian is the same as in (\ref{IntLagLFV}), and the matching given by (\ref{g1g6}) and (\ref{gLVqgRVq}) still holds.  The analysis in (\ref{gLVqgRVq})--(\ref{BKTeff2}) is valid also here.\footnote{Notice that in the Dirac model, in principle, we might have LFV BKT on the IR brane. They arise from  the RH KK neutrinos that, in analogy to the zero mode fields  (\ref{nuDKK}), contain $U_{TB}$ in their expansion.
However, this effect is indirect,  and driven by the UV mass terms $m_{\mathrm{UV},\alpha}$. 
Their impact on the model is sub-leading. We have numerically checked it in the BKT insertion approximation.} The bound on $c_l$ (\ref{clUVbound}) coming from UV uncalculable corrections also applies here.

In contrast to the Majorana model, radiative decays mediated by neutrinos and charged gauge bosons are no longer negligible, even in the absence of BKT.\footnote{The same is valid for CP violating effects, that we have not studied in the Dirac model.}
Interestingly enough, in this case we have been able to find a reasonable analytic formula for $A_R^{(W)}$, see (\ref{BRana2}),  working in the flavour basis where Yukawa couplings are treated as perturbative mass insertions.
Given the difficulty of finding such formulae, we report in the following some details on how (\ref{BRana2}) has been derived.  
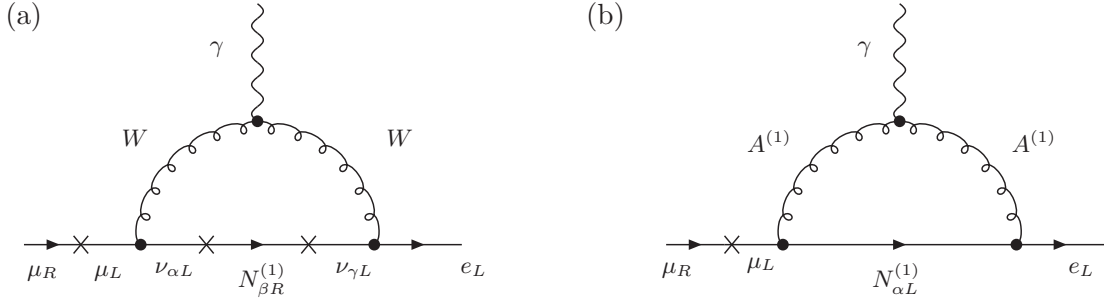
\begin{figure}[t]
\begin{center}
  \begin{picture}(400,100) (0,0)
  \setlength{\unitlength}{1.pt}
\SetScale{1.1}
    \SetColor{Black}

 \Text(0,100)[t]{(a)}

   \Text(7,1)[t]{\footnotesize{$\mu_R$}}
     \Text(32,1)[t]{\footnotesize{{$\mu_L$}}}
         \Text(57,2)[t]{\footnotesize{$\nu_{\alpha L}$}}
            \Text(90,2)[t]{\footnotesize{$N_{\beta R}^{(1)}$}}
           \Text(125,2)[t]{\footnotesize{$\nu_{
         \gamma L}$}}
              \Text(42,52)[t]{\footnotesize{$W$}}
                  \Text(142,52)[t]{\footnotesize{$W$}}
                  \Text(73,85)[t]{\footnotesize{$\gamma$}}

  \Vertex(80,50){2}
\Photon(80,50)(80,90){2}{4}

    \Text(170,2)[t]{\footnotesize{$e_L$}}

      \ArrowLine(0,8)(20,8)
        \Line(17,5)(22,11)
  \Line(22,5)(17,11)
         \Line(20,8)(40,8)
         \ArrowLine(40,8)(120,8)
     \Vertex(40,8){2}
  \Line(60,5)(65,11)
  \Line(65,5)(60,11)
    \Line(95,5)(100,11)
  \Line(100,5)(95,11)
    \Vertex(120,8){2}
       \ArrowLine(120,8)(150,8)
\GlueArc(80,8)(40,0,180){2}{12}

  \Text(220,100)[t]{(b)}

   \Text(247,2)[t]{\footnotesize{$\mu_R$}}
       \Text(279,4)[t]{\footnotesize{$\mu_L$}}
         \Text(330,2)[t]{\footnotesize{$N_{\alpha L}^{(1)}$}}
               \Text(282,52)[t]{\footnotesize{$A^{(1)}$}}
                   \Text(382,52)[t]{\footnotesize{$A^{(1)}$}}
                      \Text(318,85)[t]{\footnotesize{$\gamma$}}
  \Vertex(300,50){2}
\Photon(300,50)(300,90){2}{4}
    \Text(400,2)[t]{\footnotesize{$e_L$}}
      \ArrowLine(220,8)(240,8)
      \Line(240,8)(260,8)
          \ArrowLine(260,8)(340,8)
     \Vertex(260,8){2}
  \Line(240,5)(245,11)
  \Line(245,5)(240,11)
    \Vertex(340,8){2}
       \ArrowLine(340,8)(370,8)
\GlueArc(300,8)(40,0,180){2}{12}  
      \end{picture}
  \caption{Leading one-loop graphs contributing to $\mu\rightarrow e \gamma$ in the flavour basis for (a) $W$ and (b) $A^{(1)}$ exchange. The crosses represent Yukawa coupling insertions.} 
\label{fig:muegammaLR}
  \end{center}
  \end{figure}
We still adopt a KK approach and keep, for each 5D fermion field, only the first KK resonance. Even in this approximation, an analytic computation is complicated by the large number of fields that are present. The most important point to note is that the flavour violation comes  from the singlet fields $N_\alpha^{(1)}$ and $N_{\alpha R}^{(0)}$, arising from the expansion of the 5D fields (\ref{nuDKK}).
The zero modes  $N_{\alpha R}^{(0)}$, due to their ultra-localization towards the IR brane, are effectively decoupled and can be neglected. 
Among the massive KK gauge bosons, the leading contribution comes from the charged gauge field 
$A^{(1)}$ in the $SO(5)/SO(4)$ coset,  since it directly couples $N_\alpha^{(1)}$ to the SM leptons. We can then safely neglect the $SU(2)_L\times SU(2)_R$ massive gauge fields $W_L^{(1)}$ and $W_R^{(1)}$. 
We neglect the tiny neutrino Yukawa couplings, their only effect being the rotation of the SM neutrinos with $U_{TB}$. Charged lepton and gauge Yukawa mixing are in first approximation negligible. The relevant Lagrangian terms are the following:
\be
{\cal L}  \supset \bar N_R^{(1)} Y_{N\nu} U_{TB} l_L^{u(0)}+ 
  \frac{g}{\sqrt{2}} \bar l_L^{d(0)} \hat W U_{TB} l_L^{u(0)} + \bar N_L^{(1)} g_{N}^L \hat A^{(1)} l_L^{d(0)} + h.c.
\label{effLagnuegamma}
\ee
where the flavour index has been omitted.
The Yukawa couplings $Y_{N\nu}$  are flavour non-diagonal with roughly  the following structure: $Y_{N\nu} \simeq U_{TB}^t Y_{0,N\nu} + \delta Y_{N\nu}$, where $Y_{0,N\nu}$ is a number and $\delta Y_{N\nu}$ a matrix in flavour space with $|\delta Y_{N\nu}|\ll |Y_{0,N\nu}|$. 
The gauge couplings $g_{N}^{L}$ are also flavour violating and have the approximate form $g_{N}^{L} \simeq U_{TB}^t g_0^{L} + \delta g^{L}$, where $g_0^{L}$ is a number and $\delta g^{L}$ a matrix in flavour space with $|\delta g^{L}|\ll |g_0^{L}|$. It turns out that the leading contribution to $A_R$ comes from the first term in square brackets in $A_R^{(W)}$,  see (\ref{ALAR}). Indeed, the potential enhancement of the second term coming from the muon mass in the denominator is compensated by the smallness of the Yukawa coupling responsible for a non-vanishing RH coupling
$C^\mu_{iR}$.
The leading contributions coming from the $W$ and $A^{(1)}$ exchange are depicted in figure~\ref{fig:muegammaLR}.
Notice that no Yukawa insertion in the loop is needed in the diagram (b), because the relevant gauge interactions are already flavour violating.
 The computation of the two diagrams gives:
 \be
 A_{R}^{W}  \simeq \frac{-ie}{96\pi^2} \frac{Y_{0,N\nu}^2 \delta m_{N^{(1)}}}{m_{N^{(1)}}^3}\,, \ \ \ \ 
A_{R}^{A^{(1)}}  \simeq   \frac{-5ieg_0^L  \delta g^L}{32\sqrt{3}\pi^2 g^2}  \frac{m_{W}^2}{m_{A^{(1)}}^2}\,,  \ee
 where $\delta m_{N^{(1)}} = m_{N^{(1)}_\mu}-m_{N^{(1)}_e}$ is the mass splitting before EWSB and for simplicity we have taken $\delta g^L$ 
 all equal in flavour space. Terms proportional to $m_{N^{(1)}} \delta Y_{N\nu}$ are sub-leading and have been neglected in $A_R^{W}$.
 In $A_R^{A^{(1)}}$ we keep the leading terms in the expansion $m_{N^{(1)}}\ll m_{A^{(1)}}$. Indeed, in most of the parameter space, due to the chosen b.c. for the singlet in the ${\bf 5}$, the neutrinos $N^{(1)}_\alpha$ (which should be identified with the fields $\hat \nu_\alpha$ defined in subsection 2.2) are sensibly lighter than the $SO(5)/SO(4)$ gauge field $A^{(1)}$. In particular, for $c_l>1/2$, $m_{N^{(1)}}$ become very light. 
 Roughly speaking, it turns out that $|\delta g^L| \lesssim |\delta m_{N^{(1)}}/m_{N^{(1)}}|$ and $|Y_{0,N\nu}|\lesssim m_{W}$,
so that  $A_{R}^{A^{(1)}}/A_{R}^{W}\sim (m_{N^{(1)}}/m_{A^{(1)}})^2\ll 1$ and the dominant contribution comes from the exchange of the SM $W$ boson. Expanding (\ref{mNana}) up to ${\cal O}(|m_{{\rm UV},\alpha}|^2)$ and using the ZMA formula (\ref{massneutrinoD}),
we get the following estimate for the branching ratio: 
\be
BR(\mu\rightarrow e\gamma) \simeq \frac{3 \alpha  }{8\pi}\left|\frac{\Delta m_{{\rm sol}}^2 }{m_{W}^2}\frac{\log^{-2} \big(\frac{R^\prime}{R}\big)}{6(1+2c_l)\Gamma(c_\nu+3/2)f_{c_l}^2}
\Big(\frac{Y_{0,N\nu}}{R^\prime}\Big)^2\Big(\frac{R^\prime}{R}\Big)^{2c_\nu-1}\right|^2\,.
\label{BRana2}
\ee
The Yukawa coupling $Y_{0,N\nu}$ depends of course on the input parameters as well, but there seems to be no simple expression for it. We have checked, by comparison with the full numerical computation, that (\ref{BRana2}) is accurate at the ${\cal O}(10\%)$ level.  The branching ratio crucially depends on the values of $c_\nu$ and $c_l$.

  \begin{figure}[t!]
\begin{minipage}[t]{0.465\linewidth} 
\begin{center}
\includegraphics*[width=\textwidth]{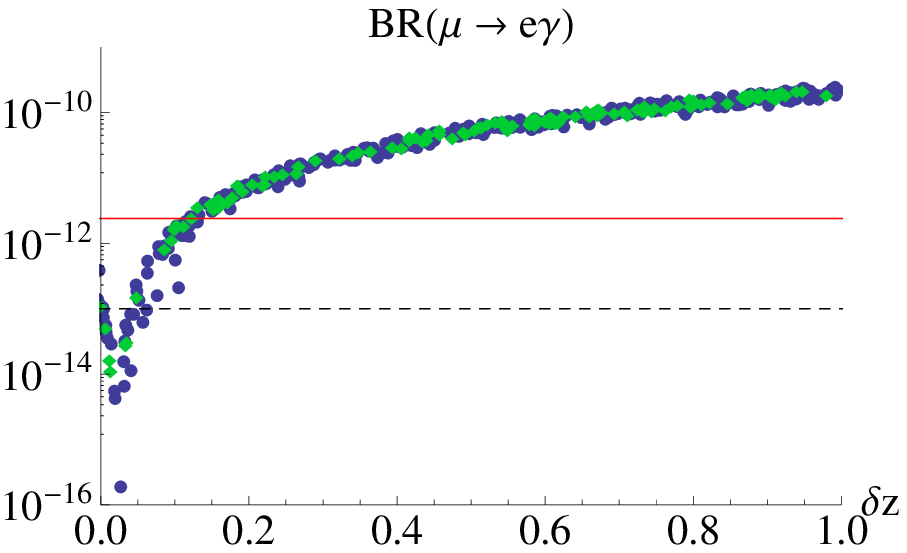}
\end{center}
\end{minipage}
\hspace{0.5cm} 
\begin{minipage}[t]{0.48\linewidth}
\begin{center}
\includegraphics*[width=\textwidth]{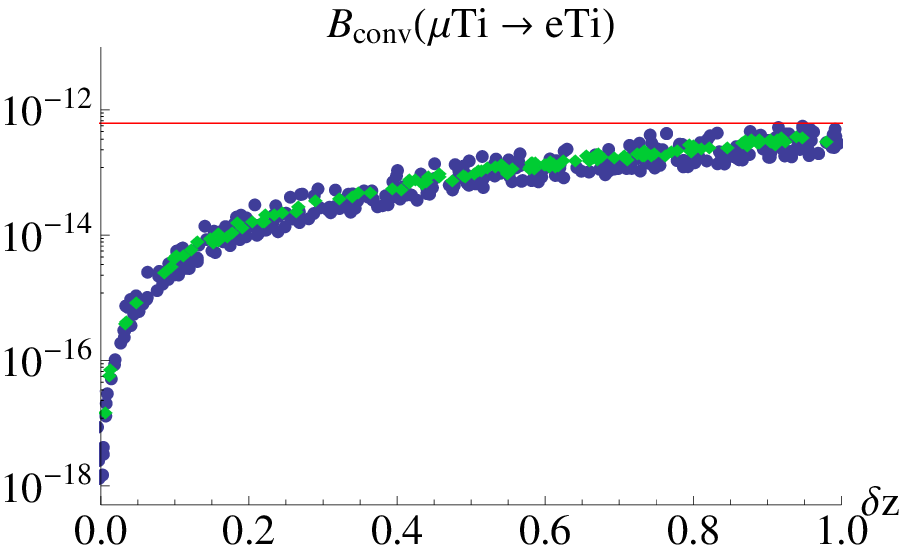}
\end{center}
\end{minipage}
\caption{\small Branching ratio of $\mu\rightarrow e \gamma$ and $\mu-e$ conversion in Ti as a function of the UV BKT $\delta z\equiv z_{\mu l}-z_{e l}$ in the Dirac model.  The continuous (red) and dashed (black) lines in the left panel represent the current  \cite{MEGactual}  and the expected future bound \cite{MEG} given by the MEG experiment. The (red) line in the right panel is the experimental bound as given by SINDRUM II \cite{Sindrum2}. The plots refer to the Dirac model with $c_l=0.52$, $c_\nu = 1.33$, $h=1/3$ and normal neutrino mass hierarchy. 
 The IR masses $m_{{\rm IR},\alpha}^l$ are random numbers chosen
between $0.05$ and $1.5$ for $m_{{\rm IR},e,\mu}^l$ and $0.5$ and $1.5$ for $m_{{\rm IR},\tau}^l$
(blue points) or all set to one (green diamonds).
 The masses $m_{\mathrm{UV},\alpha}$ are chosen such that the lightest neutrino mass is 
  $m_0=0.01$ eV and the best fit values \cite{nudata} of the solar and atmospheric mass square differences $\Delta m_{\rm{sol}}^2 = 7.59 \times 10^{-5}$ eV$^2$ and 
$\Delta m_{\rm{atm}}^2 = 2.40 \times 10^{-3}$ eV$^2$ are reproduced using (\ref{massneutrinoD}), corrected for the effect of the BKT.}
\label{fig:mue-mueconv-D}
\end{figure}
When the UV BKT in (\ref{LBKT}) are considered, the branching ratio of $\mu \rightarrow e \gamma$ 
receives extra contributions of the form (\ref{ARWana}) that for $|\delta z|\gtrsim 0.02$ dominate. 
Unfortunately, it is not simple to derive a reasonably accurate analytic expression for  $BR(\mu\rightarrow e\gamma)$ in this case. 
We plot in figure~\ref{fig:mue-mueconv-D} the branching ratio of
$\mu\rightarrow  e\gamma$ and $\mu-e$ conversion in Ti for the Dirac model for $c_l=0.52$ and $c_\nu=1.33$, setting all phases to zero. As shown in subsection \ref{subsec:devD}, the deviation of $g^\alpha_{\nu_L}$ from its SM value puts a strong bound on $h$, $h \lesssim 1/10$, see (\ref{alphabound}). We take here a value of $h=1/3$ in order to compare the results for the LFV processes in the Majorana model with those in the Dirac model.\footnote{As we mentioned, one might take the optimistic point of view that the bound (\ref{alphabound}) might be a signal of new exotic hidden physics rather than a true bound for the model. Anyhow, for $h\leq 1/10$, no bound would arise from $BR(\mu\rightarrow e\gamma)$ or $B_{conv} (\mu {\rm Ti} \to e  {\rm Ti})$, since both scale as $h^4$,
and the points in figure~\ref{fig:mue-mueconv-D} should be scaled down by two orders of magnitude.}
The bounds are mainly governed by the UV BKT, with a very mild dependence on $m^l_{\rm{IR},\alpha}$. As can be seen from figure~\ref{fig:mue-mueconv-D},  both processes depend quadratically on $\delta z$, as expected from (\ref{DLestimate}) and (\ref{ARWana}), and
the relative difference $|\delta z|$ of the UV localized BKT is constrained to be smaller than 0.15 in order to pass the actual MEG bound of $2.4 \times 10^{-12}$. This becomes smaller than $\lesssim 0.05$ to pass the  expected future MEG bound $BR(\mu\to e\gamma) < 10^{-13}$. For such small values of $\delta z$, cancellations between the contribution (\ref{BRana2}) and the one associated with the UV localized BKT can occur and further suppress $BR(\mu\to e\gamma)$, see figure~\ref{fig:mue-mueconv-D}.  The results for $BR(\mu\rightarrow 3e)$ and $B_{conv}(\mu  {\rm Ti}\rightarrow e {\rm Ti})$ are automatically below the current experimental bounds, as soon as $BR(\mu\to e\gamma)$ is below the new limit set by the MEG Collaboration. Radiative $\tau$ decays, $\tau\to\mu\gamma$ and $\tau\to e\gamma$, have branching ratios $\lesssim 10^{-9}$. 
The branching ratio of $\mu\to e\gamma$ in the Majorana model is about two orders of magnitude smaller than the one in the Dirac model, for equal values of $c_l$. In contrast, $B_{conv}(\mu  {\rm Ti}\rightarrow e {\rm Ti})$ is similar, being governed in both cases by the same tree-level FCNC.

Concerning the lepton mixing angles, we find $\sin^2 \theta_{23}$ well within the experimentally allowed $1\sigma$ range and $\sin^2 \theta_{12}$ still within the $2\sigma$ range \cite{nudata}. The value of $\sin^2 \theta_{13}$
is smaller than $10^{-8}$.  As already discussed in detail in the case of the Majorana model with non-universal masses
$m_{\rm{IR},\alpha}^\nu$, an inverted neutrino mass hierarchy is disfavoured, because the solar mixing angle receives in general too large corrections, while the atmospheric mixing angle still remains within the experimentally
allowed $1\sigma$ range and $\sin^2 \theta_{13} \lesssim 10^{-8}$. A noteworthy effect in the Dirac scenario is the rather large deviation of the lepton mixing matrix $U_{PMNS}$ from unitarity. For example, in the case of the set of
parameters used to generate the plots shown in figure~\ref{fig:mue-mueconv-D}, we checked that the diagonal elements of $U_{PMNS}^\dagger U_{PMNS}$ and $U_{PMNS} U_{PMNS}^\dagger$ can deviate up to 0.05 from one. Their off-diagonal elements 
are in general much smaller. The non-unitarity of $U_{PMNS}$ is associated with the non-decoupling of the light states $N_\alpha^{(1)}$.
This is in sharp contrast with the results found in the Majorana model in which the deviation from unitarity of $U_{PMNS}$ is in general less than $10^{-3}$ for the diagonal elements of  $U_{PMNS}^\dagger U_{PMNS}$ 
and $U_{PMNS} U_{PMNS}^\dagger$.

 \section{Conclusions} 

We have introduced a class of 4D HCHM based on the non-abelian flavour group $S_4\times \Z_3$,  where lepton masses can be naturally reproduced and nearly TB lepton mixing is predicted. Both Majorana and Dirac neutrinos can be accommodated. 
A small breaking of the flavour symmetry for charged leptons is disfavoured  in the composite sector, typically leading to a too large deviation of the coupling of the $\tau$ to the $Z$ from its SM value. The latter observation is linked to the choice of representations of the discrete flavour group used for the LH and RH charged leptons and needed to forbid large flavour violating effects. It  applies to more general  constructions based on different flavour groups. The breaking of the flavour symmetry for neutrinos in the composite sector can be large in the Dirac model, whereas it must be small in the Majorana model to suppress too large deviations from TB mixing.

We have also constructed two explicit realizations of our framework in terms of 5D gauge-Higgs unification theories. We have computed in detail
the relevant bounds coming from LFV processes in the charged lepton sector and shown that no significant constraints arise in both models.
In the Majorana model, all the spectrum of fermion resonances is above the TeV scale,\footnote{This should be contrasted with \cite{delAguila:2010vg}, where light fermion resonances appear.} while in the Dirac case light (sub--TeV) neutral fermions appear and are responsible for a too large deviation of the coupling of neutrinos to the $Z$ from its SM value. 
A particularly economic and successful Majorana model can be constructed by postulating a $\Z_2$ exchange symmetry on the IR boundary which protects neutrinos from being affected by the flavour symmetry breaking. 
In both models, Majorana and Dirac, two CP phases are present. We have not studied in detail their effects on the lepton EDM, but have argued that in the Majorana model these are expected to be negligibly small. 

Overall, the 5D Dirac model performs worse than the 5D Majorana model but, of course, this does not necessarily imply that more natural HCHM based on the Dirac scenario cannot be constructed, rather we might have missed to find a better representative. 

%%%%%%%%%%%%%%%%%%%%%%%%%%%%%%%%%%%%%%%%%%%%

\section*{Note Added}
\label{noteadded}
%%%%%%%%%%%%%%%%%%%%%%%%%%%%%%%%%%%%%%%%%%%%

During the final stages of the preparation of this paper new experimental results have been released by the T2K \cite{T2K} and MINOS \cite{MINOS}
Collaborations indicating that  $\theta_{13}=0$ is disfavoured at the level of $2.5 \sigma$ and $89\%$ confidence level, respectively. Subsequently,
three groups \cite{fogli,schwetz,maltoni} have performed a global fit of the available neutrino data finding at different levels of significance $\theta_{13}\neq 0$. The
strongest indication of $\theta_{13} \neq 0$ is found by \cite{fogli}  at a level of (more than) $3 \sigma$, while the analysis in \cite{maltoni} shows that the mixing angle $\theta_{13}$ is still
compatible with zero at the latter level. The best fit value of $\sin^2 \theta_{13}$ is $0.01 \div 0.02$ in all three analyses \cite{fogli,schwetz,maltoni}. If
such sizable value of $\sin^2 \theta_{13}$ will be confirmed in the future, our models (with Dirac and Majorana neutrinos, respectively)
become disfavoured, because they generically foresee small values of $\sin^2 \theta_{13}$ below $10^{-4}$ without additional (new) sources giving rise to $\theta_{13} \neq 0$.
However, this does not rule out HCHM with flavour symmetries (broken in the manner as proposed by us) in general, since other mixing patterns, see e.g. \cite{dATFH}, apart from TB mixing,
can be implemented as well \cite{HSfuture}.

\section*{Acknowledgments}

We would like to thank Csaba Csaki, Christophe Grojean and Jose Santiago for useful e-mail correspondence and Serguey  Petcov for useful conversations.

\appendix

%%%%%%%%%%%%%%%%%%%%%%%%%%%%%%%%%%%%%%%%%%%%%%%%%%%%%%%%%%%%%%%
\section{Group Theory of $S_4$}
\label{app:S4}
%%%%%%%%%%%%%%%%%%%%%%%%%%%%%%%%%%%%%%%%%%%%%%%%%%%%%%%%%%%%%%%

In this appendix we briefly recapitulate the group structure of $S_4$ and discuss the decomposition of the $S_4 \times \Z_3$ representations under the
subgroups $\Z_3^{(D)}$ (preserved in the composite sector/on the IR brane) and $\Z_2 \times \Z_2 \times \Z_3$ (preserved in the elementary sector/on the UV brane), respectively. 
$S_4$ is the permutation group of four distinct objects and is isomorphic to the symmetry group $O$ of a regular octahedron. It has 24 distinct elements
and five real irreducible representations: ${\bf 1}$, ${\bf
  1^\prime}$, ${\bf 2}$, ${\bf 3}$ and ${\bf 3^\prime}$, out of which only the two triplets are faithful.
We define $S_4$ with the help of three generators $S$, $T$ and $U$\footnote{$S_4$ can also be defined in terms of only two generators.}
which are of the following form for the five different representations:
\begin{center}
\begin{math}
\begin{array}{llll}
{\bf 1}:       & S=1 \; , & T=1 \; ,  & U=1 \ ,\\[2mm]
{\bf 1^\prime}:    & S=1  \; , & T=1 \; ,  & U=-1\ ,\\[2mm]
{\bf 2}: & S= \left( \begin{array}{cc}
    1&0 \\
    0&1
    \end{array} \right) \; ,
    & T= \left( \begin{array}{cc}
    \omega&0 \\
    0&\omega^2
    \end{array} \right) \; ,
    & U=  \left( \begin{array}{cc}
    0&1 \\
    1&0
    \end{array} \right)\ ,\\[2mm]
{\bf 3}: & S= \frac{1}{3} \left(\begin{array}{ccc}
    -1& 2  & 2  \\
    2  & -1  & 2 \\
    2 & 2 & -1
    \end{array}\right) \; ,
    & T= \left( \begin{array}{ccc}
    1 & 0 & 0 \\
    0 & \omega^{2} & 0 \\
    0 & 0 & \omega
    \end{array}\right) \; ,
    & U= - \left( \begin{array}{ccc}
    1 & 0 & 0 \\
    0 & 0 & 1 \\
    0 & 1 & 0
    \end{array}\right)\ ,\\[2mm]
{\bf 3^\prime}: & S= \frac{1}{3} \left(\begin{array}{ccc}
    -1& 2  & 2  \\
    2  & -1  & 2 \\
    2 & 2 & -1
    \end{array}\right) \; ,
    & T= \left( \begin{array}{ccc}
    1 & 0 & 0 \\
    0 & \omega^{2} & 0 \\
    0 & 0 & \omega
    \end{array}\right) \; ,
    & U= \left( \begin{array}{ccc}
    1 & 0 & 0 \\
    0 & 0 & 1 \\
    0 & 1 & 0
    \end{array}\right)\ ,
\end{array}
\end{math}
\end{center}
and fulfill the relations
\bea \nonumber
&& S^2 = 1 \; , \;\; T^3 = 1 \; , \;\; U^2= 1 \; ,\\[0mm]
&&  (S T)^3 = 1 \; , \;\; (S U)^2 = 1 \; , \;\; (T U)^2 = 1 \; , \;\;
  (S T U)^4 = 1\; .
\eea
Note that $S$ and $T$ alone generate the group $A_4$, and similarly, that the two generators $T$ and $U$
alone generate the group $S_3$. The character table of $S_4$ can be found in, e.g., \cite{Lomont}.
The Kronecker products are of the form
\bea\nonumber
&&\bf 1 \times {\bs \mu} = {\bs \mu} \;\; \forall \;\; {\bs \mu} \; , \;\; 1^\prime \times 1^\prime =1 \; , \;\; 1^\prime \times 2 = 2 \; ,\\[0mm]
\nonumber
&&\bf 1^\prime \times 3 = 3^\prime \; , \;\; 1^\prime \times 3^\prime = 3 \; ,\\[0mm]
\nonumber
&&\bf 2 \times 2 = 1 + 1^\prime + 2 \; , \;\; 2 \times 3 = 2 \times 3^\prime = 3 + 3^\prime\; ,\\[0mm]
&&\bf 3 \times 3 = 3^\prime \times 3^\prime = 1 + 2 + 3 + 3^\prime \; , \;\;
3 \times 3^\prime = 1^\prime + 2 + 3 + 3^\prime \; .
\eea
The Clebsch Gordan coefficients can be found in, e.g., \cite{HKL_S4}. For $( c_1 ,c_2 , c_3 )^t$, $( \tilde c_1 , \tilde c_2 , \tilde c_3 )^t \sim {\bf 3}$,
the invariant under $S_4$ is of the form $c_1 \tilde c_1 + c_2 \tilde c_3 +  c_3 \tilde c_2$.
 Note  that the choice of $T$ being complex for the
real representations ${\bf 2}$, ${\bf 3}$ and ${\bf 3'}$ leads, for 
 $(\phi_1,\phi_2)^t\sim {\bf 2}$, $(\psi_1,\psi_2,\psi_3)^t\sim {\bf 3}$ and
$(\psi'_1,\psi'_2,\psi'_3)^t\sim {\bf 3'}$, to conjugate fields transforming as
$(\phi_2^*,\phi_1^*)^t\sim {\bf 2}$,
$(\psi_1^*,\psi_3^*,\psi_2^*)^t\sim {\bf 3}$ and
$(\psi_1^{\prime *},\psi_3^{\prime *},\psi_2^{\prime *})^t\sim {\bf 3'}$.

The decomposition of the $S_4\times \Z_3$ representations under $\Z^{(D)}_3$ is given by 
\bea \nonumber
&({\bf 1}, \omega^j)& \;\;\; \rightarrow \;\;\; \omega^j \\ \nonumber
&({\bf 1^\prime}, \omega^j)& \;\;\; \rightarrow \;\;\; \omega^j \\ \nonumber
&({\bf 2}, \omega^j)& \;\;\; \rightarrow \;\;\; \omega^{j+1} + \omega^{j+2} \\ \nonumber
&({\bf 3}, \omega^j)& \;\;\; \rightarrow \;\;\; \omega^j + \omega^{j+2} + \omega^{j+1} \\ 
&({\bf 3^\prime}, \omega^j)& \;\;\; \rightarrow \;\;\; \omega^j + \omega^{j+2} + \omega^{j+1}
\eea 
for $j=0,1,2$.
Since the generator $T$ is diagonal in the group basis chosen by us, we can easily see that, e.g. for $\psi_i \sim ({\bf 3},1)$,
$\psi_1$ transforms as $1$ under $\Z^{(D)}_3$, $\psi_2$ as $\omega^2$ and $\psi_3$ as $\omega$ . 
The decomposition of the $S_4$ representations under the subgroup $\Z_2\times \Z_2$ generated by $S$ and $U$, with $S U= U S$,
is\footnote{The external $\Z_3$ factor remains unbroken in this case and is omitted in the following.}
\bea \nonumber
&{\bf 1}& \;\;\; \rightarrow \;\;\; (1,1) \\ \nonumber
&{\bf 1^\prime}& \;\;\; \rightarrow \;\;\; (1,-1) \\ \nonumber
&{\bf 2}& \;\;\; \rightarrow \;\;\; (1,1) + (1,-1) \\ \nonumber
&{\bf 3}& \;\;\; \rightarrow \;\;\; (1,-1) + (-1,1) + (-1,-1)\\ 
&{\bf 3^\prime}& \;\;\; \rightarrow \;\;\; (1,1) + (-1,1) + (-1,-1)
\eea
where $(\pm 1,\pm 1)$ indicate the transformation properties under the two $\Z_2$ factors of $\Z_2\times \Z_2$.
Since $S$ and $U$ are not diagonal, the decomposition of the $S_4$ representations under $\Z_2 \times \Z_2$ is non-trivial.
For $\phi_i \sim {\bf 2}$ we get 
\be
\frac{1}{\sqrt{2}} \left( \phi_1 \pm \phi_2 \right) \sim (1,\pm 1)\,,
\ee
for a triplet $\psi_i \sim {\bf 3}$
\be
\frac{1}{\sqrt{3}} \left( \psi_1 + \psi_2 + \psi_3 \right) \sim (1,-1) \, , \;\;
\frac{1}{\sqrt{2}} \left( \psi_2 - \psi_3 \right) \sim (-1,1)\,,  \;\;
\frac{1}{\sqrt{6}} \left( 2 \, \psi_1 - \psi_2 - \psi_3 \right) \sim (-1,-1)
\ee
and similarly for $\psi^\prime_i \sim {\bf 3'}$ under $S_4$
\be
\frac{1}{\sqrt{3}} \left( \psi^\prime_1 + \psi^\prime_2 + \psi^\prime_3 \right) \sim (1,1) \, , \;\;
\frac{1}{\sqrt{6}} \left( 2 \, \psi^\prime_1 - \psi^\prime_2 - \psi^\prime_3 \right) \sim (-1,1)\,, \;\;\; 
\frac{1}{\sqrt{2}} \left( \psi^\prime_2 - \psi^\prime_3 \right) \sim (-1,-1) \; .
\ee

%%%%%%%%%%%%%%%%%%%%%%%%%%%%%%%%%%%%%%%%%%%%%%%%%%%%%%%%%%%%%%%
\section{$SO(5)$ Generators and Representations}

\label{app:SO5}
%%%%%%%%%%%%%%%%%%%%%%%%%%%%%%%%%%%%%%%%%%%%%%%%%%%%%%%%%%%%%%%

We list here the explicit choice of $SO(5)$ generators and $SU(2)_L\times SU(2)_R$ embedding used in the paper.
Denoting by 
\be
t^{ab}_{ij} = - t^{ba}_{ij} =  \delta_i^a \delta_j^b -  \delta_i^b \delta_j^a
\ee
the 10 anti-symmetric generators of $SO(5)$, where $a,b=1,\ldots,5$ label the generators
and $i,j$ their matrix components, we take
\bea
t^1_L &= & -\frac i2 (t^{23}+t^{14}), \ \ t^2_L = -\frac i2 (t^{31}+t^{24}), \ \ t^3_L = -\frac i2 (t^{12}+t^{34}), \ \  \nn \\
t^1_R&= & -\frac i2 (t^{23}-t^{14}), \ \ t^2_R = -\frac i2 (t^{31}-t^{24}), \ \ t^3_R = -\frac i2 (t^{12}-t^{34}), \ \  \nn \\
t^{\hat a} & = &  -\frac{i}{\sqrt{2}} t^{a5}, \ \ \hat a = 1,2,3,4 \,.
\eea
In this basis, $t^{1,2,3}_L$ generate $SU(2)_L$, $t^{1,2,3}_R$ generate $SU(2)_R$ and
$t^{\hat 1,\hat 2,\hat 3,\hat 4}\in SO(5)/SO(4)$. 
A fermion multiplet $\Psi_5$ in the ${\bf 5}$ of $SO(5)$ decomposes as ${\bf 5} = ({\bf 2,2}) \oplus ({\bf 1,1})$ under $SU(2)_L\times SU(2)_R$ and can be written as follows:
\be
\Psi_5 =\frac{1}{\sqrt{2}}\left(
\begin{matrix}
i (u_+ - d_{-}) \\
- (u_+ + d_-) \\
-i (u_- + d_+) \\
 u_- - d_+ \\
\sqrt{2} n
\end{matrix}\right), \ \
\ee
where 
\be
q_\pm =\left(
\begin{matrix}
u_\pm \\
d_\pm 
\end{matrix}\right) \ \
\label{doubLRDef}
\ee
are the two doublets with $T_{3R} = \pm 1/2$, respectively, forming the bi-doublet, and $n$ is the singlet.
A fermion multiplet $\Psi_{10}$ in the ${\bf 10}$ of $SO(5)$ decomposes as ${\bf 10} = ({\bf 2,2}) \oplus ({\bf 1},{\bf 3}) \oplus ({\bf 3},{\bf 1})$ under $SU(2)_L\times SU(2)_R$ and can be written as follows:
\bea
\Psi_{10} & = &\frac{t^1_L + i t^2_L}{\sqrt{2}} \phi^+ +  \frac{t^1_L - i t^2_L}{\sqrt{2}} \phi^-+  t^3_L  \phi^0+
\frac{t^1_R + i t^2_R}{\sqrt{2}} \chi^+ +  \frac{t^1_R - i t^2_R}{\sqrt{2}} \chi^-+  t^3_R  \chi^0 \nn \\
 && -
 \frac{t^{\hat 1} + i t^{\hat 2} }{\sqrt{2}} u_+ +  \frac{t^{\hat 1}  - i t^{\hat 2} }{\sqrt{2}} d_-+ 
  \frac{t^{\hat 3}  + i t^{\hat 4} }{\sqrt{2}} u_- +  \frac{t^{\hat 3}  - i t^{\hat 4} }{\sqrt{2}} d_+\,,
\eea
where $\phi^\pm, \phi^0$ form the $SU(2)_L$ triplet ,  $\chi^\pm, \chi^0$ the $SU(2)_R$ triplet, and $u_\pm$, $d_\pm$ 
are the components of the bi-doublet, as defined in (\ref{doubLRDef}).

\section{Expressions for $A_L$ and $A_R$}

\label{app:ALAR}

We report here the explicit expressions for $A_{R/L}^{(W)}$, $A_{R/L}^{(Z)}$ and $A_{R/L}^{(H)}$, in terms of the couplings defined in (\ref{IntLagLFV}) in the main text:
\bea
A_R^{(W)}  & = & \frac{-ie}{64\pi^2}\sum_{i,V^-} \frac{m_{W}^2}{m_{V^-}^2} \Big(C_{iL}^{e *} C_{iL}^{\mu} f_W(z_i^{V^-}) + \frac{M_i}{m_\mu} C_{iL}^{e *} C_{iR}^{\mu} g_W(z_i^{V^-})\Big),
A_L^{(W)}  = A_R^{(W)}(C_{iR}^a \leftrightarrow C_{iL}^a),
\nn  \\
A_R^{(Z)}  & = &\frac{-ie}{64\pi^2}\sum_{a,V^0} \frac{m_{W}^2}{m_{V^0}^2} \Big(D_L^{ea} D_L^{a \mu} f_Z(z_a^{V^{0}}) + \frac{M_a}{m_\mu} D_L^{ea} D_{R}^{a \mu} g_Z(z_a^{V^0})\Big), 
A_L^{(Z)}  = A_R^{(Z)}(D_{R}^{a b} \leftrightarrow D_{L}^{a b}), \nn \\
 A_R^{(H)} & = &\frac{-ie}{64\pi^2}\sum_{a} \frac{m_{W}^2}{m_H^2} \Big(Y_{e a}^* Y_{\mu a} f_H(z_a^H) + \frac{M_a}{m_\mu}Y_{ea}^*   Y_{a \mu }^*  g_H(z_a^H)\Big), A_L^{(H)}  =A_R^{(H)}(Y \leftrightarrow Y^\dagger),
\label{ALAR}
\eea
with $z_i^{V^-}\!\!=M_i^2/m_{V^-}^2$,  $z_a^{V^0}\! = \! M_a^2/m_{V^0}^2$,  $z_a^H = M_a^2/m_H^2$, $m_{W}$ and $m_H$ the SM $W$ and Higgs masses, and
\bea
f_W(z) & = & \frac{1}{6(z-1)^4}\Big(10-43 z+78 z^2-49z^3+18z^3 \log z+4z^4\Big)\,, \nn \\
g_W(z) & = & \frac{1}{(z-1)^3}\Big(4-15z+12z^2-6 z^2 \log z-z^3\Big)\,,  \eea \bea
f_Z(z) & = & \frac{1}{6(z-1)^4}\Big(8-38 z+39 z^2-18z^2 \log z-14z^3+5z^4\Big)\,, \nn \\
g_Z(z) & = & \frac{1}{(z-1)^3}\Big(4-3z+6z\log z - z^3 \Big)\,, \eea \bea
f_H(z) & = & \frac{-1}{6(z-1)^4}\Big(2+3 z +6 z\log z -6 z^2+z^3\Big)\,, \nn \\
g_H(z) & = & \frac{-1}{(z-1)^3}\Big(3+2\log z -4z+z^2 \Big)\,.
\eea
All the expressions above are in agreement with the results of \cite{Lavoura:2003xp}.

%%%%%%%%%%%%%%%%%%%%%%%%%%%%%%%%%%%%%%%%%

\end{document}